\documentclass[12pt,cite,a4paper]{article}

\pdfoutput=1

\usepackage{amsmath}
\usepackage{amssymb}
\usepackage{a4wide}
\usepackage{graphicx}

\numberwithin{equation}{section}

\begin{document}

\begin{titlepage}

\rightline{MIT-CTP 4158}
\vspace{2.5truecm}

%%%%%%%%%%%%%%%%%
\centerline{\Large \bf  Warped black holes in 3D general massive gravity}

\vspace{1.5truecm}

\centerline{Erik Tonni}

\vspace{1cm}

\centerline{{\it  Center for Theoretical Physics,}}
\vspace{.1cm}
\centerline{{\it  Massachusetts Institute of Technology,}}
\vspace{.1cm}
\centerline{{\it Cambridge, MA 02139, USA}}

\vspace{.5cm}
\centerline{{\tt tonni@mit.edu}}

\vspace{2.5truecm}

%%%%%%%%%%%%%%%%%
\centerline{\bf Abstract}
\vspace{.5truecm}

We study regular spacelike warped black holes in the three dimensional general massive gravity model,  which contains both the gravitational Chern-Simons term and the linear combination of curvature squared terms characterizing the new massive gravity besides the Einstein-Hilbert term.
The parameters of the metric are found by solving a quartic equation, constrained by an inequality that imposes the absence of closed timelike curves.
Explicit expressions for the central charges are suggested by exploiting the fact that these black holes are discrete quotients of spacelike warped $AdS_3$ and a known formula for the entropy.
Previous results obtained separately  in topological massive gravity and in new massive gravity are recovered as special cases.

\vspace{2truecm}

\end{titlepage}

\section*{Introduction}

Massive gravity in three dimensional spacetime has a long story that goes back to the topological massive gravity (TMG) \cite{Deser:1982vy, Deser:1981wh}, where the gravitational Chern-Simons term (CS) is added through a coupling $1/\mu$ to the usual Einstein-Hilbert term (EH), which includes also the cosmological constant. The space $AdS_3$ with radius $\ell$ (whose isometry group is $SL(2,\mathbb{R}) \times SL(2,\mathbb{R})$) is a classical solution for every value of the coupling but 
a consistent model for quantum gravity in three dimensions with a CS term besides the EH one can be defined only at the special point $\mu \ell =1$ (the chiral point) \cite{Li:2008dq, Maloney:2009ck}.

Another model called new massive gravity (NMG) has been recently introduced \cite{Bergshoeff:2009hq, Bergshoeff:2009aq} as a consistent theory for massive gravitons in three dimensions. It is defined by adding to the EH term  curvature squared terms (we recall that in three dimensions the Riemann tensor can be expressed in terms of the Ricci tensor, therefore the only curvature squared terms one can construct are $R^2$ and $R_{\mu\nu} R^{\mu\nu}$) with a characteristic relative coefficient between them. 
The coupling $1/m^2$ multiplying this characteristic linear combination is parameterized by the mass $m$ of the graviton. This parity preserving model has been found by requiring that at the linearized level about the Minkowski vacuum it is equivalent to the Pauli-Fierz theory for a massive spin 2 field (see also \cite{Bergshoeff:2009tb}). 
A different way to find it has been proposed in \cite{Sinha:2010ai}.
In NMG the gravitons have two polarization states of helicity $\pm 2$ and mass $m$; while in TMG only a single mode of helicity $2$ and mass $\mu$ propagates. Points of the parameter space with peculiar features occur also in this model. The supersymmetric extension of this model has also been studied \cite{Andringa:2009yc,Bergshoeff:2010mf}.

Given the Lagrangians of TMG and NMG, it is natural to consider a general model which contains,
besides the ordinary EH term, both the gravitational CS term and the linear combination of curvature squared terms of the NMG. This model has been called general massive gravity (GMG) \cite{Bergshoeff:2009hq, Bergshoeff:2009aq, Liu:2009pha}. It violates parity and the $\pm 2$ helicity modes propagate with different masses.
It is interesting to look for further consequences arising from the interplay of the three terms (EH, CS and NMG) defining this model.

As mentioned above, in TMG the $AdS_3$ vacua are perturbatively unstable for  $\mu \ell \neq 1$. 
Nevertheless, in \cite{Anninos:2008fx} it has been shown that in the non chiral region and for $\mu\ell \neq 3$ there are two vacuum solutions given by warped $AdS_3$ spaces. 
The metric of the warped $AdS_3$ is non-Einstein (i.e. $R_{\mu\nu}$ is not proportional to $g_{\mu\nu}$) and this makes the analysis of these spaces more complicated. Moreover, the warping breaks the $SL(2,\mathbb{R}) \times SL(2,\mathbb{R})$ isometry group of $AdS_3$ down to $SL(2,\mathbb{R}) \times U(1)$, where the $U(1)$ isometry can be either spacelike or timelike.
Given one of these two situations, then the fiber over $AdS_2$ can be either stretched or squashed, depending on the value of the warping parameter, and this influences the causal structure of the spacetime \cite{Bengtsson:2005zj}.
In \cite{Anninos:2009zi}, by employing the consistent set of boundary conditions proposed in \cite{Compere:2008cv, Compere:2007in} (see also \cite{Compere:2009zj}), it has been shown that in TMG the spacelike warped $AdS_3$ vacua are stable. The situation is similar to one encountered for  $AdS_3$  at the chiral point, with the difference that now the region of the parameter space is larger ($\mu\ell >3$).\\
In this region, regular (namely free of closed timelike curves) black hole solutions of TMG which are asymptotic to warped $AdS_3$ with a spacelike $U(1)$ have been studied in \cite{Nutku:1993eb,Gurses:1994ab, Moussa:2003fc, Bouchareb:2007yx,Moussa:2008sj} and then in \cite{Anninos:2008fx} a coordinate transformation has been exhibited that maps the spacelike stretched black holes into the spacelike warped $AdS_3$ (stretched), showing also that the known warped black holes are discrete quotients of the spacelike warped $AdS_3$, like the BTZ black hole is a discrete quotient of $AdS_3$ \cite{Banados:1992gq,Carlip:1994gc,Maldacena:1998bw}.
The mass and the angular momentum of these warped black holes in TMG have been computed in \cite{Bouchareb:2007yx} by further developing the ADT technique proposed in \cite{Abbott:1982jh,Deser:2002rt, Deser:2002jk,Deser:2003vh}. This improvement has allowed to check the first law of thermodynamics for these spaces, since the entropy was known by using the results of \cite{Kraus:2005vz,Solodukhin:2005ah,Tachikawa:2006sz}. 
In \cite{Anninos:2008fx}, the fact that these warped black holes are discrete quotients of spacelike warped $AdS_3$ has been exploited to rewrite the entropy in a form that resembles the Cardy  formula. This result has lead to propose explicit expressions for the central charges $c_L$ and $c_R$ of the CFT dual to the TMG with suitable  asymptotically streched boundary conditions.
A strong check of this result has been done in \cite{Compere:2008cv}, where $c_R$ has been computed through an asymptotic symmetry group analysis, starting from the results of \cite{Bouchareb:2007yx}.
In \cite{Blagojevic:2009ek} both $c_L$ and $c_R$ of \cite{Anninos:2008fx} have been checked by employing the method developed in \cite{Blagojevic:2008bn} and imposing the equality between the entropy obtained through the Cardy formula and the gravitational one, in order to fix a residual ambiguity.

In NMG regular warped black holes have been obtained in  \cite{Clement:2009gq}, right after the model was proposed. In  \cite{Clement:2009gq} their entropy has been computed by the Wald formula \cite{Wald:1993nt,Jacobson:1993vj,Iyer:1994ys} while the mass and the angular momentum have not been found through a generalization of the ADT analysis for curvature square terms to non (covariantly) constant curvature backgrounds, but through the method of the super angular momentum (see also \cite{Clement:1994sb} for TMG). 
This method provides the angular momentum and the mass, but in the computation of the latter a term occurs that is not known. This term is fixed by imposing the first law of thermodynamics, which therefore cannot be used anymore as a consistency check of the procedure. As emphasized in \cite{Clement:2009gq}, a generalization of the ADT technique in NMG to non-Einstein backgrounds is still missing.

In this paper we study regular warped black holes in GMG.\\
After having introduced the GMG model and its equations of motion (section \ref{section eom}),
we discuss the two metrics we are interested in, namely the spacelike warped $AdS_3$ and the warped black hole. Then we exhibit a coordinate transformation that maps the warped black hole into the spacelike warped $AdS_3$ and slightly generalizes the result of \cite{Anninos:2008fx}, showing that the former (in the stretched regime) is a discrete quotient of the latter (section \ref{section metrics}). The parameters of the black hole characterizing the warping can be found by solving a quartic equation whose coefficients are functions of the couplings and of the cosmological constant (section \ref{algebraic relations}). Moreover, requiring the absence of closed timelike curves imposes a strong restriction to the domain within which we have to look for the solutions of the quartic equation (section \ref{section no CTC}). The warped black holes of TMG and NMG discussed above are recovered as special cases in our analysis (section \ref{section known cases}). Another known result \cite{Bergshoeff:2009hq,Bergshoeff:2009aq} about $AdS_3$ as a solution of GMG is obtained as a check (appendix \ref{AdS3 appendix}).
In GMG we find that there are regions of the parameter space where two regular warped black holes are allowed and, by exploiting the formulas for the roots of a cubic and a quartic equation (reviewed in the appendix \ref{appendix roots}), we give evidences that there are no points in the parameter space leading to three or four allowed solutions (section \ref{section GMG}).\\
In the last part of the paper (section \ref{entropy section}), the thermodynamical aspects of these warped black holes are briefly considered by providing their entropy, mass and angular momentum, which are computing by adding up known results from TMG \cite{Moussa:2003fc,Bouchareb:2007yx,Moussa:2008sj} and NMG \cite{Clement:2009gq} given in another coordinate system (see the appendix \ref{Clement coords}). The first law of thermodynamics is satisfied but this is not an independent check of the results because, as recalled above, in NMG it has been implemented in the computation of the mass \cite{Clement:2009gq}.
In order to understand some possible quantum theory underlying this model by employing the powerful tools of the AdS/CFT correspondence, we rewrite the entropy of the warped black holes found here in a form resembling the Cardy formula in 2D CFT (section \ref{section CFT}). These procedure has been employed to study the warped black hole in TMG \cite{Anninos:2008fx}, in analogy to the well established one developed for the BTZ black hole \cite{Strominger:1997eq}.
Explicit expressions for the central charges of the 2D boundary theory are therefore suggested. They are checked against the existing expressions for the same quantities proposed separately in TMG \cite{Anninos:2008fx} and NMG \cite{Kim:2009jm} (in NMG the entropy function method \cite{Sen:2005wa} has been employed), which are recovered.\\
In the appendix \ref{appendix selfdual and null} we briefly give the solutions for the self-dual warped black holes and the z-warped null black holes in GMG, which have been studied respectively in \cite{Chen:2010qm} and \cite{Gibbons:2008vi,Anninos:2010pm} for TMG.

We remark that warped spaces arise in various contexts besides the three dimensional massive gravity.
We find it important to mention, for instance, that the near horizon geometry of an extreme Kerr black hole (denoted by NHEK), a rotating black hole in four dimensions whose mass and angular momentum are related in a particular way, at fixed polar angle is given by a spacelike warped $AdS_3$ \cite{Bardeen:1999px} and this fact has represented a strong hint to the formulation of the Kerr/CFT correspondence \cite{Guica:2008mu,Bredberg:2009pv}. 
It is also important to understand the string realizations of the warped spaces considered here (see \cite{Orlando:2010ay} for recent developments in TMG).

%%%%%%%%%%%%%%%%%%%%%%%%%%%%%%%%%%%%%%%%%%

\section{General massive gravity}
\label{section eom}

In this section we introduce the action of the model and its equations of motion (e.o.m.).

The action of the general massive gravity (GMG) in three dimensions reads (we adopt the mostly plus convention for the metric)
\begin{equation}
\label{gmg action}
S[g_{\mu\nu}]\,=\,
\frac{1}{16\pi G}
\int \sqrt{-g}\left(\tilde{\sigma} R-2\lambda m^2 
+\frac{1}{\mu}\,\mathcal{L}_{\textrm{CS}}
+\frac{1}{m^2}\,\mathcal{L}_{\textrm{NMG}}\right) d^3x
\end{equation}
where the gravitational Chern-Simons (CS) term $\mathcal{L}_{\textrm{CS}}$ and
the new massive gravity (NMG) term $\mathcal{L}_{\textrm{NMG}}$  are defined respectively as follows
\begin{equation}
\mathcal{L}_{\textrm{CS}} = 
\frac{1}{2}\,\varepsilon^{\mu\nu\rho}
\left(\Gamma^\alpha_{\mu\beta} \,\partial_\nu \Gamma^\beta_{\rho\alpha}
+\frac{2}{3}\,\Gamma^\alpha_{\mu\beta} \Gamma^\beta_{\nu\gamma}\Gamma^\gamma_{\rho\alpha}\right)
\hspace{1.6cm}
\mathcal{L}_{\textrm{NMG}}
=
R_{\mu\nu} R^{\mu\nu}-\frac{3}{8}\,R^2\;.
\end{equation}
It is convenient to keep the freedom of choosing the sign of the EH term by introducing the factor $\tilde{\sigma} = \pm 1$ (in the literature $\tilde{\sigma} = +1$ is often called  ``right sign'' and  $\tilde{\sigma} = -1$ ``wrong sign''  for the EH term). The parameter $\lambda$ is dimensionless and characterizes the cosmological constant term, while $m$ and $\mu$ have the dimension of a mass and provide the couplings to the NMG term and to the CS term respectively.\\
Varying the action (\ref{gmg action}) w.r.t. the metric $g_{\mu\nu}$, we get the following equations of motion
\begin{equation}
\label{eom}
\tilde{\sigma} G_{\mu\nu}+\lambda m^2 g_{\mu\nu}
+ \frac{1}{2m^2}\,K_{\mu\nu} 
+\frac{1}{\mu}\,C_{\mu\nu}
= 0
\end{equation}
where $G_{\mu\nu}$ is the Einstein tensor, $C_{\mu\nu}$ is the Cotton tensor 
\begin{equation}
\label{cotton tensor}
C_{\mu\nu} \equiv \varepsilon_\mu^{\;\;\alpha\beta}\,
D_\alpha\bigg(R_{\beta\nu}-\frac{1}{4} R \,g_{\beta\nu}\bigg)
\end{equation}
coming from the gravitational CS term (we recall that $\varepsilon^{012}=+1/\sqrt{-g}$ for the ordered set of coordinates $\{x^0, x^1, x^2\}$) and the tensor $K_{\mu\nu}$, due to the NMG term, reads \cite{Bergshoeff:2009hq}
\begin{equation}
K_{\mu\nu} \equiv  
2 D^2 R_{\mu\nu}-\frac{1}{2}(D_\mu D_\nu R+ g_{\mu\nu} D^2 R)
-8 R_{\mu\alpha} R^\alpha_{\;\,\nu} +\frac{9}{2}R R_{\mu\nu}
+\left(3 R^{\alpha\beta} R_{\alpha\beta}-\frac{13}{8} R^2\right)
g_{\mu\nu}\;.
\end{equation}
By multiplying (\ref{eom}) by $\tilde{\sigma}$ we get
\begin{equation}
\label{eom bis}
G_{\mu\nu}+\tilde{\sigma} \lambda m^2 g_{\mu\nu}
+ \frac{\tilde{\sigma}}{2m^2}\,K_{\mu\nu} 
+\frac{\tilde{\sigma}}{\mu}\,C_{\mu\nu}
= 0\;.
\end{equation}
Moreover, if $g_{\mu\nu}= \ell^2 \tilde{g}_{\mu\nu}$ then the equations (\ref{eom bis}) can be written as follows
\begin{equation}
\label{eom rescaled}
\widetilde{G}_{\mu\nu}+\tilde{\sigma} \lambda m^2\ell^2\, \tilde{g}_{\mu\nu}
+ \frac{\tilde{\sigma}}{2m^2\ell^2}\,\widetilde{K}_{\mu\nu} 
+\frac{\tilde{\sigma}}{\mu\ell}\,\widetilde{C}_{\mu\nu}
= 0
\end{equation}
where the tilded quantities are referred to $\tilde{g}_{\mu\nu}$.\\
Let us mention that a method to solve this e.o.m. by employing the occurrence of Killing vectors fields has been studied in \cite{Gurses:2008wu,Gurses:2010sm}, but we will not use it here.

\section{Spacelike warped $AdS_3$ and warped black holes}
\label{section metrics}

In this section we give the metric of the spacelike warped $AdS_3$ and of the warped black holes that we are going to study. 
A coordinate transformation relating these two metrics is obtained and it
leads us to show that the warped black holes are discrete quotients of spacelike warped $AdS_3$.
This provides also the left and right temperatures that will be used later (section \ref{section CFT}) to get the expressions for the central charges.

The metric of $AdS_3$ can be written as the following spacelike (or timelike) fibration of the real line over $AdS_2$ in the coordinates $\{\tau, \sigma, u\}$
\begin{eqnarray}
\label{AdS3 spacelike}
ds^2 &=& \frac{\ell^2}{4}\big[-\cosh^2\sigma\,d\tau^2+d\sigma^2+(du+\sinh\sigma\,d\tau)^2\,\big]
\label{spacelike AdS3}\\
\rule{0pt}{.7cm}
&=&\frac{\ell^2}{4}\big[-(d\tau-\sinh\sigma\,du)^2+d\sigma^2+\cosh^2\sigma\,du^2\,\big]\;.
\label{timelike AdS3}
\end{eqnarray}
The range of each coordinate is $\mathbb{R}$ and the fiber coordinate is $u$ for (\ref{AdS3 spacelike}) and $\tau$ for (\ref{timelike AdS3}).
We consider the following warping of the spacelike fibration (\ref{spacelike AdS3})
\begin{equation}
\label{spacelike warped}
ds^2 \,=\, \frac{\ell^2 \chi^2}{4}\Big[-\cosh^2\sigma\,d\tau^2
+d\sigma^2+\gamma^2 (du+\sinh\sigma\,d\tau)^2 \Big]\;.
\end{equation}
The warping factor is $\gamma^2$ and for $\gamma^2 > 1$ we have the spacelike stretched $AdS_3$, while $\gamma^2 < 1$ characterizes the spacelike squashed $AdS_3$.
The warping factor $\gamma^2 \neq 1$ makes the metric (\ref{spacelike warped}) non-Einstein.
One could also introduce a warping for the timelike fibration (\ref{timelike AdS3}) as well and obtain a timelike warped $AdS_3$ solution, but we will not consider this case here (see \cite{Anninos:2008fx}, where this has been done for TMG).
The Ricci scalar of the metric (\ref{spacelike warped}) reads
\begin{equation}
\label{R warped}
R \,=\, \frac{2(\gamma^2-4)}{\ell^2\chi^2}
\end{equation}
and its determinant is $g = -\,\ell^6 \chi^6 \gamma^2  \cosh^2 \sigma /64$.

\noindent 
Now we consider the following black hole metric in ADM form coordinates $\{t,\theta, r\}$ (Schwarzschild coordinates)
\begin{equation}
\label{ADM gmg}
\frac{ds^2}{\ell^2} \,=\, 
-\,N(r)^2 dt^2+\frac{dr^2}{4R(r)^2 N(r)^2}+R^2(r) \big[\,d\theta+N^\theta(r)dt \,\big]^2
\end{equation}
where 
\begin{eqnarray}
R(r)^2 &\equiv& 
\frac{\zeta^2}{4}\,r \Big((1-\eta^2)r+\eta^2 (r_+ +r_-)+2 \eta \sqrt{r_+ r_- }\,\Big)\\
\rule{0pt}{.9cm}
N(r)^2 &\equiv& \zeta^2\eta^2\,\frac{(r-r_+)(r-r_-)}{4 R(r)^2}\\
\rule{0pt}{.9cm}
N^\theta(r) &\equiv&
|\zeta|\, \frac{r +  \eta \sqrt{r_+ r_-}}{2 R(r)^2}\;.
\end{eqnarray}
The ranges of the coordinates are $t \in (-\infty,+\infty)$, $r \in [0,+\infty)$ and $\theta \sim \theta +2\pi$.
There are two horizons which are located at $r_+$ and $r_-$, and the vacuum corresponds to $r_+=r_-=0$.\\
The form (\ref{ADM gmg}) for the metric has been employed in \cite{Anninos:2008fx} to study the warped black holes in TMG, while the introduction of the parameters $\zeta$ and $\eta^2$ is due to fact that we want to keep contact with the notation of \cite{Moussa:2003fc,Bouchareb:2007yx,Moussa:2008sj,Clement:2009gq}, where the warped black holes have been studied both in the context of TMG and of NMG by using coordinates different from the Schwarzschild ones. In the appendix \ref{Clement coords} we recall the coordinates system employed in \cite{Moussa:2003fc, Bouchareb:2007yx,Moussa:2008sj,Clement:2009gq} and its relation with the one adopted in (\ref{ADM gmg}).
The metric (\ref{ADM gmg}) can be also written as
\begin{eqnarray}
\label{schwarz bh gmg}
\frac{ds^2}{\ell^2} &=&
dt^2 +\frac{dr^2}{\zeta^2\eta^2(r-r_+)(r-r_-)}
+|\zeta| \Big(r + \eta \sqrt{r_+ r_-}\,\Big)dt d\theta\\
\rule{0pt}{.6cm}& & \hspace{5cm}
+\; \frac{\zeta^2}{4}\,r \Big((1-\eta^2)r+\eta^2 (r_+ +r_-)+2 \eta  \sqrt{ r_+ r_- }\,\Big)
d\theta^2\,.
\nonumber 
\end{eqnarray} 
The Ricci scalar for the warped black hole reads
\begin{equation}
\label{ricci bh}
R\,=\,\frac{\zeta^2(1-4\eta^2)}{2 \ell^2}
\end{equation}
and the determinant of its metric is $g = -\,\ell^6 /4$.
By relating the parameters of the warped metric (\ref{spacelike warped}) and the black hole (\ref{ADM gmg}) as follows
\begin{equation}
\label{clement subs}
\eta\,=\,\frac{1}{\gamma}
\hspace{3cm}
\zeta\,=\,\frac{2\gamma}{\chi}
\end{equation}
the Ricci scalars (\ref{R warped}) and (\ref{ricci bh}) become equal, but we will see below that the relation between these two metrics is much deeper.\\
From (\ref{schwarz bh gmg}) we observe that $\ell$ can always be absorbed introducing $\tilde{\zeta} \equiv \zeta/\ell$ and redefining the time and the radial coordinates as $t \equiv \tilde{t}/\ell$ and $r \equiv \tilde{r}/\ell^2$ respectively (the positions $r_\pm$ change accordingly to $r$).
Then, from (\ref{clement subs}) we have $\tilde{\zeta}=2\gamma/ (\ell \chi)$, i.e. only the combination $\chi \ell$ occurs in $\tilde{\zeta}$.\\ 
The warped black holes (\ref{schwarz bh gmg}) belong to a class of metrics with the following boundary behavior at large $r$ (see also \cite{Compere:2009zj})
\begin{equation}
\begin{array}{lll}
g_{tt}\,=\,\ell^2+ O(1/r) \hspace{1.8cm}& g_{tr}\,=\,O(1/r^2)  \hspace{.5cm}
& \displaystyle g_{t\theta}\,=\,\frac{\ell^2 |\zeta|}{2}\,r +O(1)  \\
\rule{0pt}{.9cm}
\displaystyle g_{rr}\,=\,\frac{\ell^2}{\zeta^2 \eta^2 \,r^2} +O(1/r^3)
& g_{r\theta}\,=\,O(1/r)
& \displaystyle g_{\theta\theta}\,=\,\frac{\ell^2 \zeta^2(1-\eta^2)}{4}\,r^2 +O(r)\;.
\end{array}
\end{equation}
Trough a slight modification of the result of \cite{Anninos:2008fx}, we get the following local coordinate transformation that relates the warped $AdS_3$ metric (\ref{AdS3 spacelike}) to the warped black hole
(\ref{ADM gmg}) 
\begin{eqnarray}
\label{tau}
\tau &=&
\arctan\left[\frac{2\sqrt{(r-r_+)(r-r_-)}}{2r-r_+-r_-}\,\sinh\left(\frac{\zeta^2\eta^2(r_+-r_-)}{4}\,\theta\right)\right]\\
\label{u}
\rule{0pt}{.8cm}
u &=&
\frac{|\zeta|\eta^2}{2}
\left\{2t+|\zeta| \left(\frac{r_+ + r_-}{2}+ \eta \sqrt{r_+ r_-}\right)\theta\right\}\\
\rule{0pt}{.6cm}
& &\hspace{5cm}
+\;
\textrm{arctanh} \left[\frac{2r-r_+-r_-}{r_+-r_-}\,\coth\left(\frac{\zeta^2\eta^2(r_+-r_-)}{4}\,\theta\right)\right]
\nonumber\\
\label{sigma}
\rule{0pt}{.9cm}
\sigma &=&
\textrm{arcsinh} \left[\frac{2\sqrt{(r-r_+)(r-r_-)}}{r_+-r_-}\,\cosh\left(\frac{\zeta^2\eta^2(r_+-r_-)}{4}\,\theta\right)\right]\;.
\end{eqnarray}
This proves that the warped $AdS_3$ and the warped black hole are locally equivalent.\\
The isometry group of $AdS_3$ is $SL(2,\mathbb{R})_L \times SL(2,\mathbb{R})_R$ and the Killing vectors for the $SL(2,\mathbb{R})_L$ isometries are 
\begin{eqnarray}
J_1 &=&
2\,\frac{\sinh u}{\cosh \sigma} \,\partial_\tau
- 2\cosh u  \,\partial_\sigma
+ 2 \tanh \sigma \sinh u\,\partial_u\\
\rule{0pt}{.6cm}
\label{J2}
J_2 &=& 2\partial_u\\
\rule{0pt}{.6cm}
J_3 &=& -\,2\,\frac{\cosh u}{\cosh \sigma} \,\partial_\tau
+ 2\sinh u  \,\partial_\sigma
- 2 \tanh \sigma \cosh u\,\partial_u
\end{eqnarray}
while the ones for $SL(2,\mathbb{R})_R$ read
\begin{eqnarray}
\tilde{J}_1 &=&
2 \sin \tau \tanh \sigma \,\partial_\tau
- 2\cos \tau  \,\partial_\sigma
+ 2 \,\frac{\sin\tau}{\cosh \sigma}\,\partial_u\\
\rule{0pt}{.6cm}
\label{J2 tilded}
\tilde{J}_2 &=& -\,2\cos \tau \tanh \sigma \,\partial_\tau
- 2\sin \tau  \,\partial_\sigma
- 2 \,\frac{\cos\tau}{\cosh \sigma}\,\partial_u\\
\rule{0pt}{.6cm}
\tilde{J}_3 &=& 2\partial_\tau\;.
\end{eqnarray}
They satisfy the $SL(2,\mathbb{R})$ algebra, i.e. $[J_1 , J_2]=2J_0$, $[J_0 , J_1]=-2J_2$ and $[J_0 , J_2]=2J_1$ and the same commutation relations hold among the tilded ones.
The spacelike warping  breaks the $SL(2,\mathbb{R})_L \times SL(2,\mathbb{R})_R$ isometry group of $AdS_3$ down to $SL(2,\mathbb{R})_R \times U(1)_L$, which is the isometry group of the metric (\ref{spacelike warped}): it preserves the three Killing vectors of $SL(2,\mathbb{R})_R$ and only $\tilde{J}_2$ among the ones of $SL(2,\mathbb{R})_L$.\\
Since the warped black holes are locally equivalent to the spacelike warped $AdS_3$, we can find a discrete subgroup of the $SL(2,\mathbb{R})_R \times U(1)_L$ isometries under which we have to identify the points of the spacelike warped $AdS_3$ in order to get the warped black hole, as done in \cite{Anninos:2008fx} for the warped black hole in TMG and in \cite{Banados:1992gq,Carlip:1994gc} for the BTZ black hole, which is a quotient of $AdS_3$.\\
A Killing vector $\xi$ allows to define a one parameter subgroup of the isometry group used to identify the points $P$ of the warped $AdS_3$ space as $P \sim e^{2\pi i \xi} P$ ($k=0,1,2, \dots$) \cite{Banados:1992gq}. To get the black hole through the coordinate transformation above we have to identify the points along the direction given by $\partial_\theta$ in such a way that $\theta \sim \theta +2\pi$.\\
From the explicit expressions (\ref{J2}) and (\ref{J2 tilded}) of $J_2 \in U(1)_L$ and $\tilde{J}_2  \in SL(2,\mathbb{R})_R$ respectively and from the coordinate transformations (\ref{tau}), (\ref{u}) and (\ref{sigma}), one can write the $2\pi \xi= \partial_\theta$ Killing vector of the warped black hole for $r>r_+>r_-$ as
\begin{equation}
\label{theta Killing vector}
\partial_\theta\,=\,
\pi\ell \big(T_L J_2 - T_R \tilde{J}_2\big)
\end{equation}
where the following left and right moving temperatures have been introduced
\begin{equation}
\label{TL and TR}
T_L \,\equiv\,
\frac{\eta^2 \zeta^2}{8\pi \ell}
\left(r_+ + r_- +2 \eta \sqrt{r_+ r_-}\right)
\hspace{2cm}
T_R \,\equiv\,
\frac{\eta^2 \zeta^2}{8\pi \ell}\,(r_+ - r_-)
\end{equation}
as done in \cite{Anninos:2008fx} for the warped black hole in TMG and in \cite{Maldacena:1998bw} for the BTZ black hole.
These temperatures will be employed in the section \ref{section CFT} to find the expressions for the central charges $c_L$ and $c_R$.\\
From (\ref{J2}) and (\ref{J2 tilded}), one gets that the norm of the Killing vector $\partial_\theta$ (computed through the metric (\ref{spacelike warped})) is given by
\begin{equation}
\big|T_L J_2 - T_R \tilde{J}_2\big|^2\,=\,
\ell^2\chi^2\Big\{
T_R^2\big[1+(\gamma^2-1)\cos^2\tau \cosh^2\sigma \big]
+2\gamma^2 T_R T_L \cos\tau \cosh\sigma
+\gamma^2 T_L^2
\Big\}
\end{equation}
up to a factor $\pi^2\ell^2$. If $\gamma^2 <1$ (squashed case) this norm is negative at the boundary and the quotient along the $\partial_\theta$ direction would lead to closed timelike curves \cite{Anninos:2008fx, Bengtsson:2005zj, Banados:1992gq}.
The spacelike stretched ($\gamma^2 >1$) black hole (\ref{ADM gmg}) is therefore a quotient of the spacelike stretched $AdS_3$ by a shift of $2\pi$, generated by the isometry (\ref{theta Killing vector}).

\section{Algebraic relations from the e.o.m.}
\label{algebraic relations}

Having briefly discussed the metrics, in this section we plug them into the e.o.m. (\ref{eom bis}) and find the relations between the parameters of the metrics and the ones of the Lagrangian.

As for the parameters occurring in the Lagrangian, we find it convenient to introduce
\begin{equation}
\label{0 parameters}
\lambda_0 \equiv \tilde{\sigma} \lambda m^2\ell^2
\hspace{1.5cm}
\alpha_0 \equiv  \frac{\tilde{\sigma}}{m^2 \ell^2}
\hspace{1.5cm}
\beta_0 \equiv  \frac{\tilde{\sigma}}{\mu \ell}
\end{equation}
which are the coefficients of the different terms in (\ref{eom rescaled}).
Substituting the metric (\ref{spacelike warped}) (for the warped $AdS_3$ metric also the relations (\ref{clement subs}) must be used) and the warped black hole metric (\ref{ADM gmg}) into the equations of motion (\ref{eom bis}), we find that they are satisfied provided that the following algebraic relations hold
\begin{equation}
\label{3 eqs system}
\hspace{-.2cm}
\left\{
\begin{array}{l}
\hspace{-.1cm}
16 \alpha_0\,\zeta^4 \eta^4  -120 \alpha_0\, \zeta^4 \eta^2 +64\beta_0 \,|\zeta|^3 \eta^2
-64\,\zeta^2\eta^2 +105 \alpha_0\,\zeta^4- 64 \beta_0\,|\zeta|^3
+48\,\zeta^2-64\lambda_0
 = 0 \\
\rule{0pt}{.7cm}
\hspace{-.1cm}
16 \alpha_0\,\zeta^4 \eta^4  -80 \alpha_0\, \zeta^4 \eta^2 +32 \beta_0 \,|\zeta|^3 \eta^2
 +63 \alpha_0\,\zeta^4- 32 \beta_0\,|\zeta|^3 +16\,\zeta^2+64\lambda_0
 = 0 \\
\rule{0pt}{.7cm}
\hspace{-.1cm}
4 \alpha_0\,\zeta^2 \eta^2  -21 \alpha_0\, \zeta^2 +12 \beta_0 \,|\zeta| -8
 =0\;.
\end{array}\right.
\end{equation}
It is important to remark that in the derivation of these equations we have assumed $(\zeta,\eta)\neq (0,0)$ and $\eta^2 \neq 1$. This means that the limiting case of $AdS_3$, which corresponds to $\eta^2=1$ and $|\zeta|=2$, cannot be considered directly from (\ref{3 eqs system}).
In the appendix \ref{AdS3 appendix} we give more details about the derivation of (\ref{3 eqs system}) for the spacelike warped $AdS_3$ metric, showing also how $AdS_3$ is obtained as a special case and recovering in this way a result of \cite{Bergshoeff:2009hq,Bergshoeff:2009aq}.\\ 
If $\alpha_0 \neq 0$, we can express $\eta^2$ in terms of $\zeta$ from the last equation of (\ref{3 eqs system}) as follows
\begin{equation}
\label{eta2 soln}
\eta^2
\,=\,\frac{21}{4}-\frac{3\beta_0}{\alpha_0 |\zeta|}+\frac{2}{\alpha_0 \zeta^2}\;.
\end{equation}
Then, plugging (\ref{eta2 soln}) into the first two equations of (\ref{3 eqs system}), we find that they reduce to the same equation
\begin{equation}
\label{eq zeta}
21 \alpha_0^2\,\zeta^4  -32\alpha_0\beta_0 \,|\zeta|^3
+12(4\alpha_0+\beta_0^2)\zeta^2-32\beta_0 \,|\zeta|
+16(1+\alpha_0\lambda_0)
=0
\end{equation}
which is the main algebraic equation we have to deal with.
Since our problem is invariant for $\zeta \rightarrow -\zeta$, we can restrict the analysis to $\zeta>0$ hereafter without loss of generality. In the following we will denote by $P_4(\zeta)$ the l.h.s. of (\ref{eq zeta}).
By using the definitions (\ref{0 parameters}) and $\tilde{\zeta} = \zeta/\ell$, it is useful to rewrite (\ref{eta2 soln}) in terms of the parameters of the Lagrangian (\ref{gmg action})
\begin{equation}
\label{eta2 soln phys}
\eta^2\,=\,\frac{21}{4}-\frac{3m^2}{\mu \, \tilde{\zeta}}+\frac{2\tilde{\sigma} \,m^2}{\tilde{\zeta}^2}
\end{equation}
and similarly for (\ref{eq zeta}), obtaining
\begin{equation}
\label{eq zeta phys}
\frac{21}{m^4}\,\tilde{\zeta}^4 -\frac{32}{m^2\mu}\,\tilde{\zeta}^3
+12\left(\frac{4\tilde{\sigma}}{m^2}+\frac{1}{\mu^2}\right)\tilde{\zeta}^2
-\frac{32\tilde{\sigma}}{\mu}\,\tilde{\zeta}
+16(1+\lambda)
=0\;.
\end{equation}
The relations (\ref{eta2 soln}) and (\ref{eq zeta}), or equivalently (\ref{eta2 soln phys}) and (\ref{eq zeta phys}), are crucial results for our discussions. They are the equations to study in order to find warped solutions of GMG.

\section{Absence of closed timelike curves}
\label{section no CTC}

Before studying the roots of the quartic equation (\ref{eq zeta}), in this section we consider the restriction imposed by requiring the absence of closed timelike curves \cite{Moussa:2008sj,Clement:2009gq}, which involves only the relation (\ref{eta2 soln}).

\begin{figure}[h]
\vspace{.5cm}
\begin{tabular}{ccc}
\hspace{-.3cm}
\includegraphics[width=7.5cm]{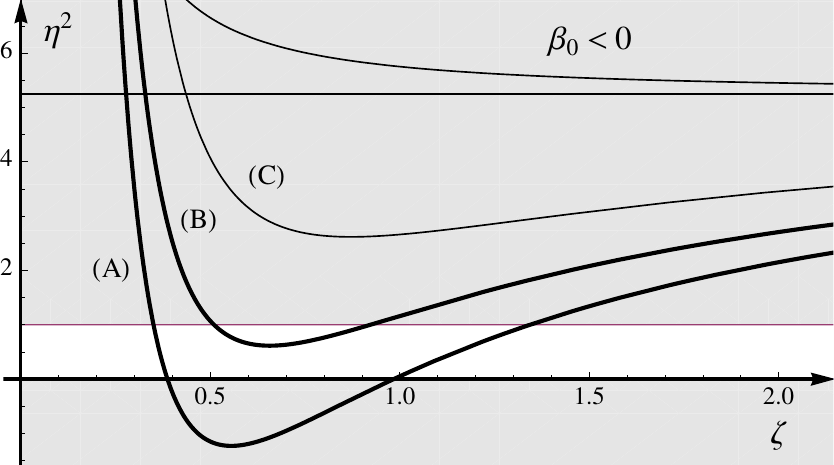}
& & 
\includegraphics[width=7.5cm]{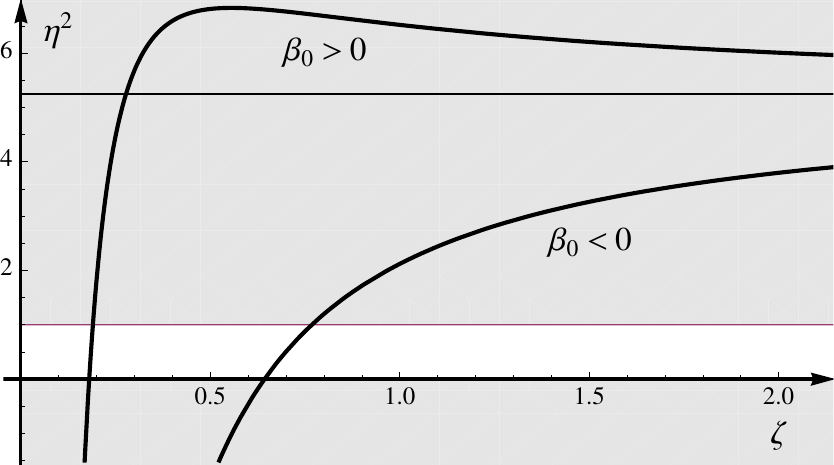}
\end{tabular}
\caption{plots of $\eta^2(\zeta)$ from (\ref{eta2 soln}). In the picture on the left the curves have $\alpha_0 >0$ while on the right $\alpha_0 <0$. The grey region denotes the regime of $\eta^2$ excluded by the condition (\ref{no CTC condition}), while the white strip represents the allowed range for $\eta^2$ which determines  on the horizontal axis the domain for the allowed values of $\zeta$. The thick curves are the ones leading to a non empty domain of allowed values for $\zeta$, while for the values of $(\alpha_0,\beta_0)$ characterizing the thin ones regular warped black holes do not exist (the curves (A), (B) and (C) has the same $\alpha_0=1$). 
\label{eta2 picture}}
\end{figure}

In the general situation $\alpha_0 \neq 0$ and $\beta_0 \neq 0$ and we must be careful because not all the real roots of (\ref{eq zeta}) for given values of the parameters $(\lambda_0,\alpha_0, \beta_0)$ lead to allowed solutions for warped black holes. Indeed, in \cite{Moussa:2008sj} it has been shown that the necessary condition for the existence of causally regular warped black holes, namely free of naked closed timelike curves, is the following
\begin{equation}
\label{no CTC condition}
0<\eta^2<1
\end{equation}
i.e. $\gamma^2 > 1$, which corresponds to the stretched case. The limiting cases of warped black holes with $\eta^2=0$ and $\eta^2=1$ must be treated separately (see \cite{Moussa:2008sj}) and they will not be considered here.\\
The expression (\ref{eta2 soln}) for $\eta^2(\zeta)$ depends on $(\alpha_0,\beta_0)$ and not on 
$\lambda_0$, therefore, through the condition (\ref{no CTC condition}), these two parameters determine a domain for $\zeta$ within which we have to look for possible roots of the quartic equation (\ref{eq zeta}). All the possible configurations of $\eta^2(\zeta)$ for $\zeta >0$ are shown in the figure \ref{eta2 picture}.

\noindent 
It is straightforward to observe that  $\eta^2(\zeta) \rightarrow \textrm{sign}(\alpha_0) \infty$ when $\zeta \rightarrow 0^+$ and $\eta^2(\zeta) \rightarrow 21/4$ when $\zeta \rightarrow +\infty$. 
Moreover, there is only one stationary point $\zeta_{\textrm{m}}$. 
The value $\zeta_{\textrm{m}}$ and $\eta^2(\zeta_{\textrm{m}})$ allow us to discriminate among the different configurations. They read
\begin{equation}
\zeta_{\textrm{m}} \,=\,\frac{4}{3}\,\beta_0
\hspace{2cm}
\eta^2(\zeta_{\textrm{m}})\,=\,
\frac{21}{4}-\frac{9\beta_0^2}{8\alpha_0}\;.
\end{equation}
It is therefore evident then when $\beta_0 <0$ the function $\eta^2(\zeta) $ has no stationary point for $\zeta>0$. In order to give the domain for the values of $\zeta$ allowed by (\ref{no CTC condition}), we have to consider the roots of $\eta^2(\zeta)=0$ and $\eta^2(\zeta)=1$, which are given respectively by  
\begin{equation}
\label{zeta10pm}
\zeta_{0,\pm} =
\frac{2}{21 \alpha_0}
\left(3\beta_0 \pm  
\sqrt{9\beta_0^2-42\alpha_0}\,\right)
\hspace{1cm}
\zeta_{1,\pm} =
\frac{2}{17 \alpha_0}
\left(3\beta_0 \pm  
\sqrt{9\beta_0^2-34\alpha_0}\,\right)\;.
\end{equation}
In the case of $\alpha_0 <0$ (see the plot on the right in the figure \ref{eta2 picture}), we have either $\eta^2(\zeta_{\textrm{m}})>21/4$ (when $\beta_0>0$) or $\zeta_{\textrm{m}}<0$ (when $\beta_0<0$), and we can always find a non empty domain of allowed values of $\zeta$ made by a single interval whose extrema are fixed by $(\alpha_0,\beta_0)$. In particular, it is given by $\{\zeta_{0,-} , \zeta_{1,-}\}$, where the analytic expressions of the extrema can be read from (\ref{zeta10pm}).\\
The case of $\alpha_0 >0$ is more subtle because there are regions of the parameter space $(\alpha_0,\beta_0)$ where $\eta^2(\zeta)$ violates (\ref{no CTC condition}) for all $\zeta>0$ and therefore causally regular warped black holes do not exist there (see the thin lines in the plot on the left in the figure \ref{eta2 picture}). For instance, this happens when $\beta_0 <0$, independently of the value of $\alpha_0$. Instead, if $\alpha_0 >0$ and $\beta_0 >0$ then we have three different situations that are characterized by $\eta^2(\zeta_{\textrm{m}})$ as follows
\begin{equation}
\label{ABC cases}
\alpha_0 >0 \hspace{.7cm} \beta_0 >0
\hspace{.5cm}\Longrightarrow \hspace{.5cm}
\left\{\begin{array}{lcl}
\eta^2(\zeta_{\textrm{m}}) <0 & \hspace{.4cm} (A)\hspace{.4cm} & 
\{\zeta_{1,-}, \zeta_{0,-}\} \,\cup\, \{\zeta_{0,+}, \zeta_{1,+}\}
\\
0<\eta^2(\zeta_{\textrm{m}}) <1 &  (B)& 
\{\zeta_{1,-}, \zeta_{1,+}\} \\
\eta^2(\zeta_{\textrm{m}}) >1 &  (C)& 
\; \emptyset\;.
\end{array}
\right.
\end{equation} 
Thus, for $\alpha_0 >0$ we can have either two disjoint intervals or one single interval of allowed values of $\zeta>0$, depending on the value of $\eta^2(\zeta_{\textrm{m}})$.\\
Once the domain of the allowed values of $\zeta>0$ has been found, in order to find a causally regular warped black hole, we have to look for the roots of the quartic equation (\ref{eq zeta}) belonging to such domain.

\section{Solutions}

In this section we study the explicit solutions of the system (\ref{3 eqs system}) constrained by the condition (\ref{no CTC condition}) imposing the absence of closed timelike curves.
First, in the section \ref{section known cases}, we check that we can recover the known warped black hole solutions in the limiting cases of TMG \cite{Anninos:2008fx, Moussa:2003fc,Bouchareb:2007yx, Moussa:2008sj} and NMG \cite{Clement:2009gq}. Then, in the section \ref{section GMG}, we consider the general case of GMG finding new warped black hole solutions.

\subsection{Known special cases: TMG and NMG}
\label{section known cases}

{\bf TMG limit.}
The TMG case is obtained by setting $\alpha_0=0$ in (\ref{3 eqs system}). 
The last equation of the system then gives 
\begin{equation}
\label{zeta TMG}
\zeta^2\,=\,\frac{4}{9\beta_0^2}
\end{equation}
and plugging this expression into the other two equations one gets
\begin{equation}
\label{eta TMG}
\eta^2\,=\,\frac{1-27\beta_0^2 \lambda_0}{4}
\end{equation}
which is the solution given in \cite{Bouchareb:2007yx} (see their eq. (4.3), where we recall that $\Lambda_{\textrm{there}}=\lambda m^2$). 
Notice that in order to recover the results of  \cite{Anninos:2008fx} in terms of their parameterization, one has to set $\lambda_0=-1$ and $\beta_0^2=1/(9\nu^2)$, which lead to $\zeta^2 = 4\nu^2$ and $\eta^2=(\nu^2+3)/(4\nu^2)$ (to match exactly with the formulas of \cite{Anninos:2008fx}, set $\zeta =2\nu$ and $\eta=-\sqrt{(\nu^2+3)/(4\nu^2)}<0$).\\

\noindent 
{\bf NMG limit.}
To recover the results in NMG obtained in \cite{Clement:2009gq}, we have to keep $\alpha_0 \neq 0$ and set $\beta_0=0$. 
The relation (\ref{eta2 soln}) simplifies to
\begin{equation}
\label{eta2 nmg}
\eta^2\,=\,\frac{21}{4}+\frac{2}{\alpha_0 \zeta^2}
\end{equation}
which coincides with the second formula in eq. (3.19) of \cite{Clement:2009gq}, provided that the notation  here is adapted to the notation there by setting $\beta_{\textrm{there}}=\eta$, $m^2_{\textrm{there}}=-m^2$, $\Lambda_{\textrm{there}}=\lambda m^2$, $\ell=1$ and $\tilde{\sigma}=+1$, which lead e.g. to $\alpha_0 \lambda_0=\lambda= - \Lambda_{\textrm{there}}/m^2_{\textrm{there}}$.
From the expression (\ref{eta2 nmg}) it is evident that the condition (\ref{no CTC condition}) can be satisfied only for $\alpha_0 <0$ (indeed, in the notation of \cite{Clement:2009gq}, $\alpha_0 = -1/m^2_{\textrm{there}}$).\\
The equation (\ref{eq zeta}) greatly simplifies for $\beta_0=0$ because the odd powers of $\zeta$ vanish and we are left with an algebraic equation of the second order in terms of $\zeta^2$ (sometimes called biquadratic equation), whose roots are given by
\begin{equation}
\label{zeta nmg}
\zeta^2 \,=\, \frac{4}{21 \alpha_0}
\Big[-6 \pm  
%\textrm{sign}(\alpha_0) 
\sqrt{3(5-7\alpha_0\lambda_0)}\,\Big]
\end{equation}
which is the solution found in \cite{Clement:2009gq} (see its eq. (3.12)).\\
In order to understand the role of  the relative position of the curves $\eta^2(\zeta)$ and $P_4(\zeta)$ for the GMG case, we find it helpful to describe it in some detail for the simpler case of NMG, where the analytic expressions are easier to deal with. 
\begin{figure}[h]
\vspace{.5cm}
\begin{center}
\includegraphics[width=10cm]{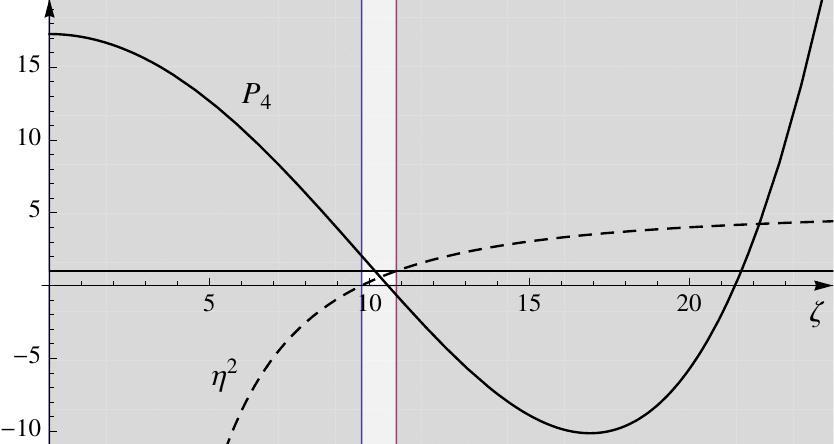}
\end{center}
\caption{NMG regime. Plot of $\eta^2(\zeta)$ from (\ref{eta2 nmg}) (dashed line) and $P_4(\zeta)\big|_{\beta_0=0}$ from (\ref{eq zeta}) with parameters set to $(\alpha_0,\lambda_0)=(-0.004,-20)$. The vertical strip is the domain $\{\zeta_{0,-} , \zeta_{1,-} \}$ determined by the condition (\ref{no CTC condition}). For the range (\ref{nmg lambda domain}) of the cosmological parameter there is a unique positive root belonging to this domain (the smaller one) and it characterizes the warped black hole solution of NMG.
\label{plot NMG root}}
\end{figure}

\noindent 
As discussed in the section \ref{section no CTC}, the condition (\ref{no CTC condition}) with $\eta^2$ given by (\ref{eta2 nmg}) fixes the interval $\{\zeta_{0,-} , \zeta_{1,-} \}$ within which we have to find $\zeta$ in order to get a causally regular black hole:
\begin{equation}
\label{nmg allowed zetas}
\{\zeta_{0,-} , \zeta_{1,-} \}\,=\,
\big\{ \sqrt{-8/(21\alpha_0)}\, , \sqrt{-8/(17\alpha_0)}\,\big\}\;.
\end{equation}
As remarked in \cite{Clement:2009gq}, in this case only one among the two positive roots of (\ref{eq zeta}) could be acceptable because it easy to see that $P_4(\zeta)$ has its minimum for $\zeta>0$ at $\sqrt{-8/(7\alpha_0)}$, which is greater than $\zeta_{1,-}$. This shows that in NMG there is at most one solution. Thus, the condition (\ref{no CTC condition}) selects one of the two positive roots of $P_4(\zeta)|_{\beta_0=0}=0$ and in particular the smaller one in (\ref{zeta nmg}), 
which corresponds to the plus sign because $\alpha_0 <0$.\\
Notice that, in order to have an allowed solution, such root must fall into the domain (\ref{nmg allowed zetas}) and this is related to the value of $\lambda_0$. Indeed, since $\lambda_0$ occurs only in the constant term of $P_4(\zeta)$, by modifying $\lambda_0$ we change only the vertical position of the curve (see the figure \ref{plot NMG root}).
The constant term of $P_4(\zeta)$ depends linearly on $\lambda_0$  once $\alpha_0$ has been fixed (this is true also when $\beta_0 \neq 0$), and therefore it is straightforward to show that imposing the existence of a solution leads to the following domain for the cosmological constant \cite{Clement:2009gq}
\begin{equation}
\label{nmg lambda domain}
-\frac{1}{21}<\lambda< \frac{35}{289}
\end{equation}
(we have used that $\alpha_0\lambda_0=\lambda$) within which there is a unique warped black hole in NMG.

\subsection{GMG solutions}
\label{section GMG}

In this section we discuss the solutions for the warped black holes in the case of GMG, i.e. when both $\alpha_0 \neq 0$ and $\beta_0 \neq 0$.  This means that we look for the roots of the quartic equation (\ref{eq zeta}) such that the condition (\ref{no CTC condition}) is satisfied.
\begin{figure}[h]
\vspace{.5cm}
\begin{center}
\includegraphics[width=10cm]{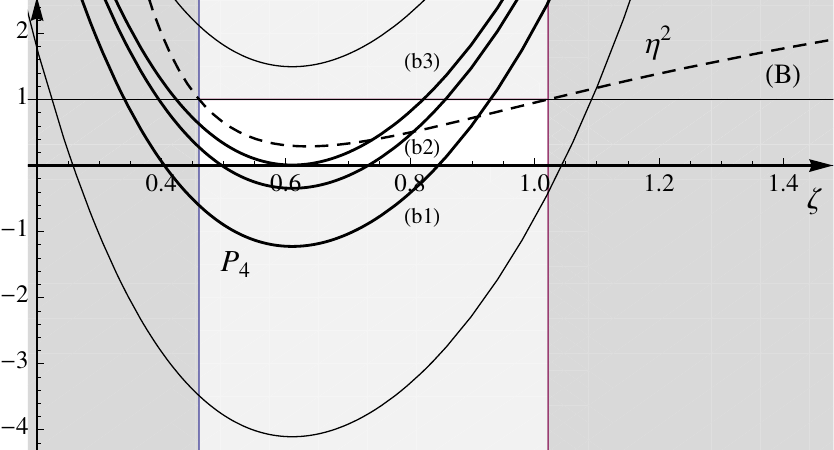}
\end{center}
\caption{warped black holes in GMG for the case denoted by (B) in (\ref{ABC cases}) and in the picture on the left in the figure \ref{eta2 picture}. The specific values are $(\alpha_0,\beta_0)=(1,2.1)$ (i.e. $k\simeq 2.42$ in the parameterization (\ref{kappa parameter})). The dashed curve is $\eta^2(\zeta)$ and it determines the allowed domain for $\zeta$ (white region). The remaining curves represent $P_4(\zeta)$ for increasing values of $\lambda_0$ going from the bottom to the top (in particular $\lambda_0 \in \{-0.27, -0.09, -0.035,-0.013,0.08\}$). Thick curves have zeros in the allowed domain for $\zeta$. For the curve (b1) and (b3) there is one acceptable solution for $\zeta$, while for the (b2) case there are two of them. 
\label{plot GMG root B}}
\end{figure}

\begin{figure}
\vspace{.5cm}
\begin{center}
\includegraphics[width=10cm]{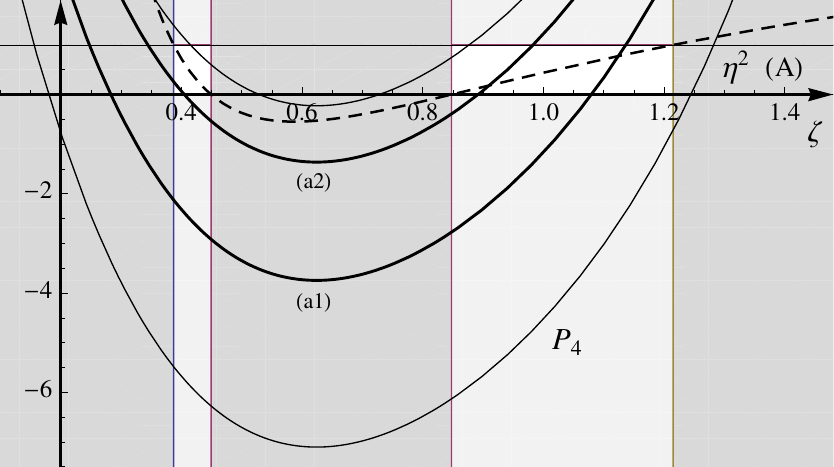}
\end{center}
\caption{warped black holes in GMG for the case (A) in the notation of (\ref{ABC cases}) and of the figure \ref{eta2 picture} (picture on the left). The plot is done for $(\alpha_0,\beta_0)=(1,2.27)$ (i.e. $k\simeq 2.07$ from (\ref{kappa parameter})). 
The dashed curve $\eta^2(\zeta)$ defines a domain for the values of $\zeta$ fulfilling the condition (\ref{no CTC condition}) made by two intervals. By increasing the cosmological parameter $\lambda_0$ through the sequence $\{-0.38, -0.17, -0.021,0.05\}$,  the curve $P_4(\zeta)$ moves upward determining different situations: either one solution (a1), or two solutions (a2) or no solutions at all (thin curves). Notice that, in this case, the value $\lambda=0$ provides a case of (a2) type.
\label{plot GMG root A}}
\end{figure}

\noindent 
We find it helpful to describe the situation qualitatively before giving any explicit expression for the roots. For any set of values for $(\lambda_0, \alpha_0, \beta_0)$ such that $(\alpha_0, \beta_0)\neq (0,0)$, we consider first the pair $(\alpha_0, \beta_0)$ because it fixes $\eta^2(\zeta)$ completely, telling us which is the situation  we are dealing with among the ones shown in the figure \ref{eta2 picture}. In particular we can decide whether there is a non empty domain of allowed values for $\zeta$ (made by either one or two intervals) or there are no values at all of $\zeta$ fulfilling the condition (\ref{no CTC condition}).\\
Once we have established the domain for the values of $\zeta$ satisfying (\ref{no CTC condition}), we have to find all the roots of the quartic equation (\ref{eq zeta}) and see whether some of them fall into such domain. At this step the value of $\lambda_0$ becomes crucial. As already observed in the section \ref{section known cases}, $\lambda_0$ occurs only in the constant term of $P_4(\zeta)$, therefore it can just modify the position of  $P_4(\zeta)$ along the vertical axis. The constant term of $P_4(\zeta)$ (i.e. $16(1+\lambda)$) depends linearly on $\lambda$ and therefore we can quite easily determine (graphically) the domain of $\lambda$ such that at least one zero of $P_4(\zeta)$ belongs to the domain of the allowed values of $\zeta$; similarly to what it has been done for NMG to get (\ref{nmg lambda domain}).\\
The new feature of GMG with respect to the TMG and NMG cases is that, for some regions of the parameter space $(\alpha_0,\beta_0)$, there are values of $\lambda$ which lead to two allowed solutions for positive $\zeta$.
In the figures \ref{plot GMG root B} and \ref{plot GMG root A} we have chosen some values of $(\alpha_0,\beta_0)$ in order to show this feature in the cases denoted by (B) and (A) respectively in the notation of (\ref{ABC cases}) and of the picture on the left of the figure \ref{eta2 picture}: two warped black holes exist in the cases of (b2) and (a2) type. \\
At this point it is natural to ask whether there are ranges for the parameters $(\lambda_0, \alpha_0, \beta_0)$ such that three of the roots or maybe all of them (four) fall into the domain of $\zeta$ allowed by (\ref{no CTC condition}). 
Unfortunately we do not have a full analytic argument to answer this question, but through the following analysis we can exclude the possibility that four allowed solutions exist and we give also some numerical evidence that the case of three allowed solutions cannot be found as well.

\begin{figure}[h]
\vspace{.5cm}
\begin{tabular}{ccc}
\hspace{-.3cm}
\includegraphics[width=7.5cm]{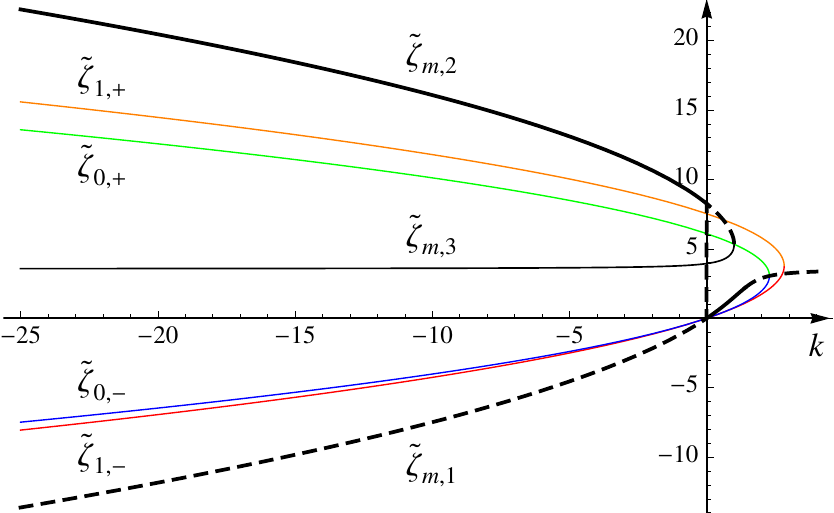}
& & 
\includegraphics[width=7.5cm]{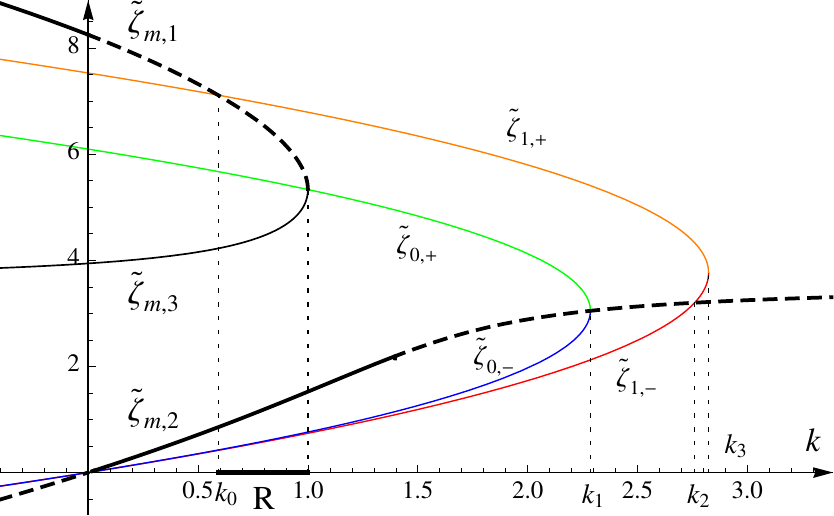}
\end{tabular}
\caption{plots of the tilded quantities introduced in (\ref{tilded zetas}) as functions of $k$ ($\beta_0>0$). The black curves are $\tilde{\zeta}_{\textrm{m},1}$ (dashed), $\tilde{\zeta}_{\textrm{m},2}$ (thick) and $\tilde{\zeta}_{\textrm{m},3}$ (thin), while the thin colored ones are $\tilde{\zeta}_{0,\pm}$ and $\tilde{\zeta}_{1,\pm}$. Three allowed solutions at most could occur only in the region R given by $k_0 < k<1$.
When $\beta_0 <0$ the thick black line and the dashed one are interchanged.
\label{stat points picture}}
\end{figure}

\noindent
In order to study the number of real zeros of $P_4(\zeta)$, it is useful to consider its stationary points, i.e. the roots of $\partial_\zeta P_4(\zeta) = 0$, which is a cubic equation. The properties of $\partial_\zeta P_4(\zeta) = 0$ are independent of the cosmological parameter $\lambda$ because it occurs only in the constant term of $P_4(\zeta)$.
The basic notions about the cubic equation that we will employ here have been collected in the appendix \ref{app cubic eq} and can be found in standard algebra textbooks (see e.g. \cite{Birkhoff1997,Irving2004}).\\
The quantity (\ref{Delta3 def}) for the cubic equation $\partial_\zeta P_4(\zeta) = 0$ reads
\begin{equation}
\Delta_3 \,=\,
-\,221184\, \alpha_0^2 \big(42\alpha_0^2-12\beta_0^2\alpha_0+\beta_0^4\big)
\big[32\alpha_0-3\beta_0^2\big]
\end{equation}
and its positivity determines the number of distinct stationary points of $P_4(\zeta)$.
Since the expression in the round brackets is always positive, the sign of $\Delta_3$ is given by the expression in the square brackets. Explicitly, for $\alpha_0 < (3/32)\beta_0^2$ we have $\Delta_3>0$ and our quartic function $P_4(\zeta)$ has three real and distinct stationary points $\zeta_{\textrm{m},j}$ ($j=1,2,3$). In particular, this is true  for $\alpha_0<0$.\\
We find it useful to introduce the following parameterization
\begin{equation}
\label{kappa parameter}
\alpha_0 \,\equiv\,
\frac{3}{32}\,k \beta_0^2
\hspace{2.4cm} k \neq 0\;.
\end{equation}
Notice that requiring $k \neq 0$ implies that we cannot recover the NMG regime, which corresponds to $\alpha_0 \neq 0$ and $\beta_0=0$.
Then, for $\beta_0 \neq 0$, we define
\begin{equation}
\label{tilded zetas}
\zeta_{0,\pm} \,\equiv\, \frac{\tilde{\zeta}_{0,\pm}}{\beta_0 k}
\hspace{1.5cm} 
\zeta_{1,\pm} \,\equiv\, \frac{\tilde{\zeta}_{1,\pm}}{\beta_0 k}
\hspace{2.5cm}
\zeta_{\textrm{m},j} \,\equiv\, \frac{\tilde{\zeta}_{\textrm{m},j}}{\beta_0 k}
\hspace{1cm} 
j= 1,2,3
\end{equation}
where $\zeta_{0,\pm}$ and $\zeta_{1,\pm}$ are given by (\ref{zeta10pm}) while $\zeta_{\textrm{m},j}$ ($j=1,2,3$) are the stationary points of $P_4(\zeta)$, which can be found by applying the formulas of the appendix \ref{app cubic eq} to the cubic equation $\partial_\zeta P_4(\zeta) = 0$.
The interesting feature of the tilded quantities is that they depend only on $k$.\\
Let us consider first the case $\beta_0 >0$, which is more interesting.
Through (\ref{rr1}), (\ref{rr2}) and (\ref{rr3}) we can find $\zeta_{\textrm{m},j}$ ($j=1,2,3$) respectively. 
In the figure \ref{stat points picture} we plot the tilded functions introduced in (\ref{tilded zetas}). 
The black curves represent $\tilde{\zeta}_{\textrm{m},j}$ and in particular the dashed one gives (\ref{rr1}), the thick one (\ref{rr2}) and the thin one (\ref{rr3}). 
By looking at these plots we can take some conclusions about the existence of three or four allowed roots.
We stress that this analysis is based on the key property that $\lambda$ occurs only in the constant term of $P_4(\zeta)$.

\noindent 
As imposed by the parameterization (\ref{kappa parameter}), only one real minimum occurs for $k>1$ (we recall that the coefficient of  $\zeta^4$ in $P_4(\zeta)$ is strictly positive, thus if there is only one stationary point then it must be a minimum). The minima of  $P_4(\zeta)$ are $\zeta_{\textrm{m},1}$ and $\zeta_{\textrm{m},2}$ and if at least one of them belongs to the interior of the domain of the values of $\zeta$ allowed by (\ref{no CTC condition}), then it is always possible to find a range for $\lambda_0$ such that two regular warped black holes exist for any value of $\lambda_0$ in that range.
The same consideration could be done for the maximum $\zeta_{\textrm{m},3}$, but it always fall outside the domain allowed by (\ref{no CTC condition}) (see the figure \ref{stat points picture}).
Having a minimum in the allowed domain is not a necessary condition to get two allowed solutions. Indeed for $1<k<k_1$ we can find examples of this situation, where the minimum does not belong to the domain of the allowed values for $\zeta$ but there exist values of $\lambda_0$ such that two zeros of $P_4(\zeta)$ are allowed (figure \ref{plot GMG root A}, case (a2)). \\
The points $k_3\simeq 2.8235$, $k_2\simeq 2.76$ and $k_1\simeq 2.2857$ are defined respectively as the solutions of $\eta^2(\zeta_{\textrm{m}})=1$, $\tilde{\zeta}_{\textrm{m},1}=\tilde{\zeta}_{0,-}$ and $\tilde{\zeta}_{\textrm{m},1}=\tilde{\zeta}_{1,-}$.
We find it curious that $k_1$ is also the solution of $\eta^2(\zeta_{\textrm{m}})=0$.
By considering $\partial_\zeta^2 P_4(\zeta) =0$, it is straightforward to see that $P_4(\zeta)$ changes its concavity twice if $\alpha_0 < (11/84)\beta_0^2$. While passing through 
this critical value (corresponding to $k \simeq 1.3968$) by varying $k$ influences the shape of $P_4(\zeta)$, it does not change the number of its zeros, which remains equal to two.
\\
A necessary condition (not sufficient) to have four regular warped black holes for some choice of the $(\alpha_0,\beta_0,\lambda_0)$ is that at least two stationary points of $P_4(\zeta)$ fall into the domain of the allowed $\zeta$. From the plots in the figure \ref{stat points picture} it is clear that this never happens; indeed the maximum $\zeta_{\textrm{m},3}$ always stays outside the domain of the allowed $\zeta$ and there is no value of $k$ such that both the minima $\zeta_{\textrm{m},1}$ and $\zeta_{\textrm{m},2}$ fall into it.
Thus, a choice of $(\alpha_0,\beta_0,\lambda_0)$ providing four regular warped black holes does not exist.
As for the existence of three roots allowed by (\ref{no CTC condition}), it is necessary to have three real stationary points and at least one of them must be in the allowed domain. 
From the figure \ref{stat points picture} we can see that in our case this happens only the region denoted by R, where $k_0 <k<1$, being $k_0$ the solution of $\tilde{\zeta}_{\textrm{m},1}=\tilde{\zeta}_{1,+}$ ($k_0 \simeq 0.5926$). Only in the region R we could find a set of values for $(\alpha_0,\beta_0,\lambda_0)$ leading to three regular warped black holes. \\
Let us consider now the case of $\beta_0<0$. From the figure \ref{eta2 picture} and the discussion of the section \ref{section no CTC} we recall that a non empty domain for the allowed values of $\zeta>0$ occurs only for $\alpha_0 <0$, i.e. $k<0$ if we employ the parameterization (\ref{kappa parameter}).
Such domain is given by the single interval $\{\zeta_{0,-} , \zeta_{1,-}\}$.
Then, by using (\ref{tilded zetas}) also in this case, we find plots like the ones shown in the figure \ref{stat points picture} but with the black thick line and the dashed one interchanged. Thus, for $\beta_0<0$ we have to consider only the picture on the left of the figure \ref{stat points picture} and ignore the region $k>0$. By considerations similar to the ones just exposed, it is clear that we cannot find three regular warped black holes for $\beta_0 <0$.

A more refined analysis is needed to give an answer to the question of the existence of three allowed zeros and the analytic expressions of the roots of the quartic equation $P_4(\zeta)=0$ given in (\ref{eq zeta}) must be employed.

\begin{figure}[h]
\vspace{.5cm}
\begin{tabular}{ccc}
\hspace{-.3cm}
\includegraphics[width=7.5cm]{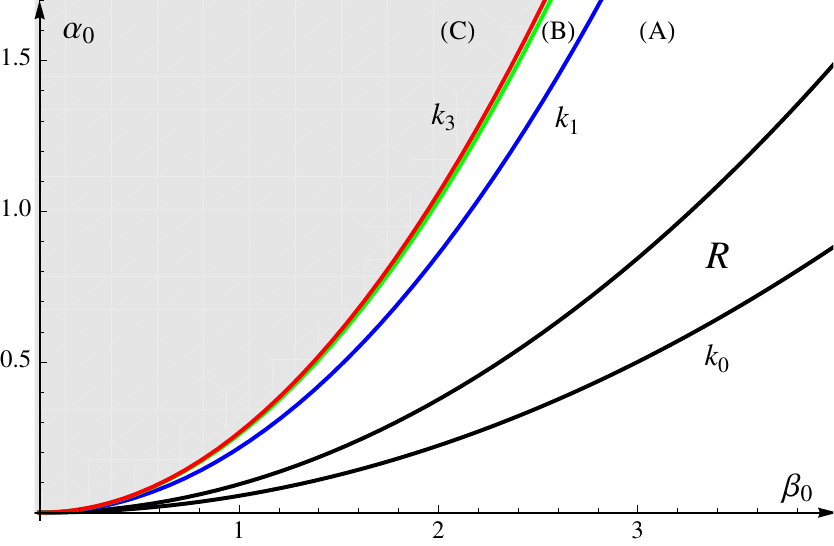}
& & 
\includegraphics[width=7.5cm]{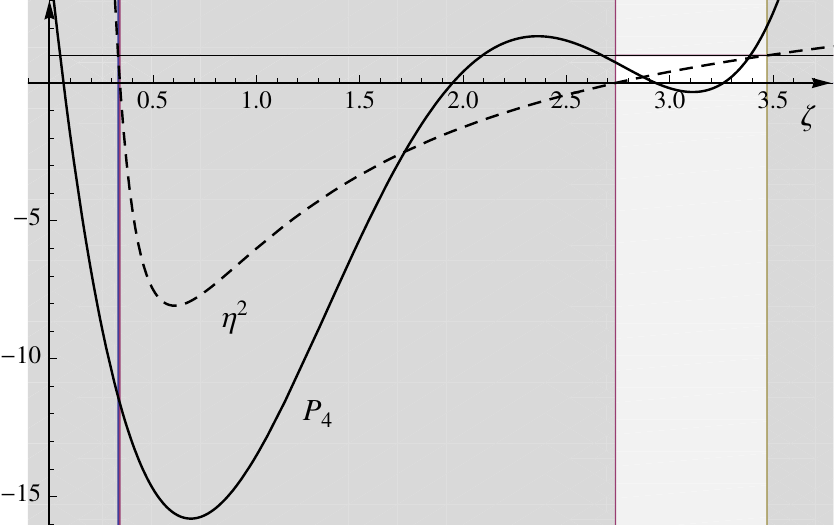}
\end{tabular}
\caption{on the left, the parameter space $(\alpha_0,\beta_0)$, where different regions are separated by particular values of $k$ (see (\ref{kappa parameter})). The region R is delimited by the black lines. In the regions (A), (B) and (C) the position of $\eta^2(\zeta)$ w.r.t. the limits imposed by (\ref{no CTC condition}) is the one shown in the figure \ref{eta2 picture} with the same notation. The green line corresponds to $k_2$ but we recall that the (B) region is for $k_1<k<k_3$.
On the right, an example of the $\eta^2(\zeta)$ (dashed curve) and $P_4(\zeta)$ in the region R ($k = 0.9$ and $(\beta_0,\lambda_0)=(2.2, -1.77)$). Notice also the tiny domain (not vanishing) of $\{\zeta_{1,-}, \zeta_{0,-}\}$; indeed we are in the case (A) of (\ref{ABC cases}) (see figure \ref{eta2 picture}).
\label{plot abregion}}
\end{figure}
\noindent 
The method to find the roots of a quartic equation is briefly reviewed in the appendix \ref{app quartic eq}. 
From the discussion reported there, we find it convenient to introduce the following parameters 
\begin{eqnarray}
\label{a GMG}
a &=& 4\left( \frac{4}{7 \alpha_0}-\frac{11 \beta_0^2}{147 \alpha_0^2}\right)
\\
\rule{0pt}{.7cm}
\label{b GMG}
b &=& 32
\left(\frac{\beta_0}{147 \alpha_0^2}-\frac{2\beta_0^3}{9261 \alpha_0^3}\right)
\\
\rule{0pt}{.7cm}
\label{c GMG}
c &=& 16\left( 
\frac{1}{21\alpha_0^2}-\frac{16 \beta_0^2}{1029 \alpha_0^3}
+\frac{80 \beta_0^4}{64827 \alpha_0^4}+\frac{\lambda_0}{21 \alpha_0}
\right)
\end{eqnarray}
which are the coefficients of the depressed quartic equation coming from (\ref{eq zeta}) (see (\ref{abc def})). Notice that the cosmological parameter occurs only in $c$.\\
In general, finding the roots of a quartic equation involves the solution of a cubic one.

In some special cases this is avoided and we can find the roots of the quartic equation by solving a couple of quadratic equations. For instance, when $\alpha_0 \lambda_0 = 5/16$ our equation (\ref{eq zeta}) becomes a quasi symmetric one (see (\ref{quasi symmetric eq}) with $\omega=1/\alpha_0$) and its solutions are 
\begin{equation}
\zeta\,=\,\frac{h\pm_{(2)}\sqrt{h^2-4/\alpha_0}}{2}
\end{equation}
being $h$ the solutions of a quadratic equation ($h$ is the r.h.s. of (\ref{soln quasi sym}))
\begin{equation}
h\,=\, \frac{1}{21 \alpha_0}
\left[ 16 \beta_0 \pm_{(1)}  
%\textrm{sign}(\alpha_0) 
\sqrt{2(2\beta_0^2-63\alpha_0)}\,\right]\;.
\end{equation}
We recall that $\pm_{(1)} $ and $\pm_{(2)} $ are independent, thus we have four distinct solutions.\\
Another situation where we do not have to solve a cubic equation is when $b=0$ (see eq. (\ref{b GMG})), i.e. $63 \alpha_0 = 2\beta_0^2$, which necessarily imposes $\alpha_0 >0$.  In this case the depressed quartic equation (\ref{dep quartic eq})
reduces to a biquadratic equation and the solutions read
\begin{equation}
\zeta\,=\,\frac{8\beta_0}{21\alpha_0}\pm_{(1)}
\sqrt{\frac{-\,a\pm_{(2)}\sqrt{a^2-4c}}{2}}
\end{equation}
where $a$ and $c$ are given by (\ref{a GMG}) and (\ref{c GMG}) respectively.

In the generic case, following the steps reported in the appendix \ref{app quartic eq}, we find that the roots of the quartic equation (\ref{eq zeta}) are given by
\begin{equation}
\label{zeta soln GMG}
\zeta\,=\,\frac{8\beta_0}{21\alpha_0}+ \frac{1}{2}\left[
\pm_{(1)}\sqrt{a+2y}
\pm_{(2)}
\sqrt{-\left(3a+2y\pm_{(1)}\frac{2b}{\sqrt{a+2y}}\right)}\,
\,\right]
\end{equation}
where $y$ is any root of the following cubic equation 
\begin{equation}
\label{eq3 y}
y^3+\frac{5}{2}\, a y^2+(2a^2-c) y +\frac{1}{2}\left(a^3-a c-\frac{b^2}{4}\right)=\,0\;.
\end{equation}
The derivation of (\ref{zeta soln GMG}) and (\ref{eq3 y}) is briefly discussed in the appendix \ref{app quartic eq}, with some additional comments about them.

\begin{figure}[h]
\vspace{.5cm}
\begin{center}
\includegraphics[width=10cm]{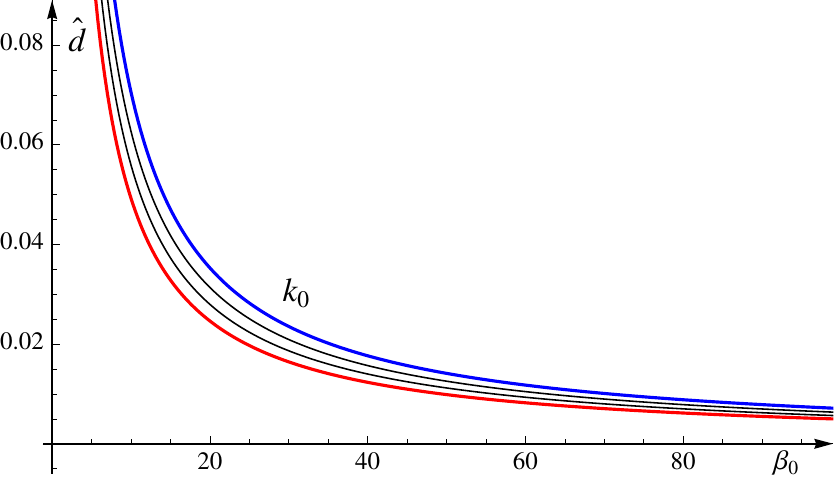}
\end{center}
\caption{plot of the difference $\hat{d}$ as function of $\beta_0$ for some values of $k$ in the region R (see figure \ref{stat points picture}, plot on the right). The blue curve corresponds to $k=k_0$, the red one to $k=1$ and then the curves for a couple of intermediate values ($k\simeq 0.8126$ and $k\simeq 0.9126$) are shown.
\label{plot no 3rd root}}
\label{no3roots picture}
\end{figure}
To close this section we exploit the formula (\ref{zeta soln GMG}) to consider the existence of three zeros of (\ref{eq zeta}) allowed by (\ref{no CTC condition}). As discussed above, they can occur only the region R (i.e. $k_0<k<1$), where the typical situation is the one shown in the figure \ref{plot abregion} (plot on the right) and it is strongly related to the value of $\lambda_0$.
From this plot, it is evident that we have to consider the distance between $\zeta_{1,-}$ and the smallest one among the four real zeros of $P_4(\zeta)$, that we will denote by $\zeta_0$. Indeed, the zero between $\zeta_{\textrm{m},2}$ and $\zeta_{\textrm{m},3}$ is not allowed because both these stationary points fall outside the allowed domain in the region R (see the figure \ref{stat points picture}, plot on the right).
Since the constant term of $P_4(\zeta)$, i.e. $16+(3/2)k\beta_0^2 \lambda_0$ (we have used (\ref{kappa parameter})), depends linearly on $\lambda_0$ through a positive coefficient, by decreasing $\lambda_0$ the positive number $\zeta_{1,-}-\zeta_0$ increases (we recall that $\zeta_{1,-}$ is indepent of $\lambda_0$). For fixed values of $(k,\beta_0)$ the distance $\zeta_{1,-}-\zeta_0$ is minimum when $P_4(\zeta_{\textrm{m},1})=0$, i.e. the limiting case where we have only one (double) solution in the set $\{\zeta_{0,+},\zeta_{1,+}\}$ of the allowed domain. We denote by $\hat{\lambda}_0=
\hat{\lambda}_0(k,\beta_0)$ the value of $\lambda_0$ corresponding to this configuration and by $\hat{d} \equiv (\zeta_{1,-}-\zeta_0)\big|_{\lambda_0=\hat{\lambda}_0}$ the difference of interest at such value.
Notice that for $k=k_0$ we find $\zeta_0\big|_{k=k_0, \lambda_0=\hat{\lambda}_0}=0$, namely $16+(3/2)k_0\beta_0^2 \hat{\lambda}_0=0$ for any value of $\beta_0$.
In the figure \ref{no3roots picture} we plot the difference $\hat{d}$ as a function of $\beta_0$ for some values of $k$ in the region R. Since this difference is always positive in the regions we have explored, we believe that there is no choice of $(\alpha_0,\beta_0,\lambda_0)$ leading to three regular warped black holes, namely three real zeros of $P_4(\zeta)$ allowed by the condition (\ref{no CTC condition}), but we do not have an analytic proof of this claim.

\section{Entropy, mass and angular momentum}
\label{entropy section}

In this section we first discuss the thermodynamics of the warped black holes found above by giving for them all the quantities occurring in the first law.
Then, in the section \ref{section CFT} we rewrite the entropy in a Cardy-like form that leads us to suggest the central charges of the two dimensional CFT on the boundary.

The temperature and angular velocity of the horizon can be read from the metric in the ADM form (\ref{ADM gmg}) respectively as follows
\begin{eqnarray}
T & \equiv & \frac{1}{2\pi}\,\sqrt{g^{rr}} \,\partial_r N\big|_{r=r_+} 
\;=\;
\frac{\eta^2 \zeta\, ( r_+ - r_- )}{4\pi \ell (r_+ + \eta \sqrt{r_+ r_-}\,)}
\\
\rule{0pt}{.7cm}
\Omega & \equiv &
 \frac{1}{\ell}\, N^\theta \big|_{r=r_+} 
 \;=\;
\frac{2}{ \ell \zeta (r_+ + \eta \sqrt{r_+ r_-}\,)}
\end{eqnarray}
where we have used that $\zeta>0$ and $\eta^2<1$. The procedure of \cite{Kraus:2005vz,Solodukhin:2005ah,Tachikawa:2006sz} to compute the entropy, which extends the Wald treatment by including the gravitational Chern-Simons term, is linear; therefore the entropy of GMG is simply given by  $S=S_{\textrm{EH}}+ S_{\textrm{CS}}+S_{\textrm{NMG}}$, namely the sum of the contributions from the three terms in the Lagrangian.
In a similar way one can compute the mass and the angular momentum.
The entropy, the mass and the angular momentum of the warped black holes have been separately computed in \cite{Moussa:2003fc,Bouchareb:2007yx,Moussa:2008sj} for TMG and in \cite{Clement:2009gq} for NMG. 
In the appendix \ref{Clement coords} we give these quantities for the warped black hole in GMG in the coordinate system used in \cite{Moussa:2003fc,Bouchareb:2007yx,Moussa:2008sj,Clement:2009gq} and we provide the change of coordinates (see (\ref{coords change}) and (\ref{rho0 omega})) between their system of coordinates and the Schwarzschild one employed in (\ref{ADM gmg}) and throughout the paper.
From (\ref{S clement}) and (\ref{SMJ change}) we find
\begin{equation}
\label{Ss warped}
S\,=\,\frac{\pi\ell\,\zeta}{4G}
\left(\frac{2}{\bar{m}^2}-\frac{1}{3}+\frac{\eta^2( r_+ - r_- )}{3\big(r_+ + \eta \sqrt{r_+ r_-}\,\big)}\right)
\big(r_+ + \eta \sqrt{ r_+ r_-}\,\big)
\end{equation}
where $\bar{m}^2 \equiv m^2/\zeta^2$ has been introduced in \cite{Clement:2009gq}. 
From (\ref{M clement}), (\ref{J clement}) and (\ref{SMJ change}) we get
\begin{eqnarray}
\label{Ms warped}
M&=& \frac{\zeta^2 \eta^2}{16 G} \left(\frac{2}{\bar{m}^2}-\frac{1}{3}\right)
\big(r_+ +r_- + 2\eta \sqrt{ r_+ r_-}\,\big)
\\
\rule{0pt}{.8cm}
\label{Js warped}
J &=& \frac{\ell \,\zeta^3 \eta^2}{16 G} 
\left[\left(\frac{1}{\bar{m}^2}-\frac{1}{6}\right)
\Big((1+\eta^2)r_+ r_- + (r_+ + r_-)\eta \sqrt{r_+ r_-}\,\Big)
-\frac{\eta^2(r_+ - r_-)^2}{12}
\right]\,.\hspace{1cm}
\end{eqnarray}
Given these expressions, we can verify that the first law of thermodynamics
\begin{equation}
\label{1st law}
d M\,=\,T dS+\Omega dJ
\end{equation}
is satisfied for independent variations of  $r_+$ and $r_-$.\\
Notice that (\ref{1st law}) is not a consistency check of the expressions (\ref{Ss warped}), (\ref{Ms warped}) and (\ref{Js warped}) because, as remarked in \cite{Clement:2009gq}, in NMG the computation of the mass has been done through the super angular momentum method and at some point an unknown constant occurs which has been fixed by imposing the validity of the first law. Thus, the first law of thermodynamics is an ingredient to get the mass in NMG. Instead, in TMG the mass and the angular momentum have been computed in \cite{Bouchareb:2007yx} through an extension of the ADT method \cite{Abbott:1982jh,Deser:2002rt, Deser:2002jk,Deser:2003vh}; therefore the first law provides a true consistency check of the results. 
In the case of GMG, since we are mixing the results from TMG and NMG, the first law is already implemented in the computation of the mass because of the method employed in NMG.

\subsection{CFT considerations}
\label{section CFT}

Given the expression (\ref{Ss warped}) for the entropy and the temperatures $T_L$ and $T_R$ introduced in (\ref{TL and TR}), we can rewrite $S$ in the following suggestive way
\begin{equation}
\label{S cardy}
S \,=\, \frac{\pi\ell}{3}\big(c_L T_L + c_R T_R\big)
\end{equation}
as done in \cite{Anninos:2008fx} for the warped black holes in TMG. The expressions of $c_L$ and $c_R$ are defined through (\ref{S cardy})  and  in our case they read
\begin{equation}
\label{cL and cR}
c_L \,=\, \frac{3\ell}{G\,\eta^2\zeta}\left(\frac{2}{\bar{m}^2}-\frac{1}{3}\right)
\hspace{1.6cm}
c_R \,=\, \frac{3\ell}{G\,\eta^2\zeta}\left(\frac{2}{\bar{m}^2}+\frac{2\eta^2-1}{3}\right)
\end{equation}
where $\bar{m}^2$ has been introduced after (\ref{Ss warped}). 
It is non trivial that (\ref{cL and cR}) are independent of $r_\pm$.\\ 
Moreover, we always have $c_R > c_L$, indeed their difference
\begin{equation}
\label{cL minus cR}
c_L-c_R \,=\,
-\frac{2\ell}{G \zeta}
\end{equation}
is negative because $\zeta >0$ (we assume positive $G$ and $\ell$).\\
%*****REMOVE****\\
%A simple consistency check of (\ref{cL minus cR}) comes from the observation that, from (\ref{tilded zetas}), if $\beta_0 \rightarrow 0$ (for fixed $k \neq 0$) then the extrema of the domain identified by (\ref{no CTC condition}) $\zeta_{1,\pm}$ and $\zeta_{0,\pm}$ diverge and this implies that also $\zeta \rightarrow +\infty$ as well in (\ref{cL minus cR}), providing the expected result that $c_L = c_R$ in the EH regime (we recall that $\beta_0 \rightarrow 0$ with fixed $k$ means $\alpha_0 \rightarrow 0$ from (\ref{kappa parameter}), i.e. we cannot recover the NMG regime with the parameterization (\ref{kappa parameter})).\\
%***********************\\
Notice that we face the puzzling situation that $c_L - c_R$ does not depend only on the coupling of the CS term, as expected from the general analysis of \cite{Kraus:2005vz,Kraus:2005zm}, but a very complicated dependence on all the parameters of the Lagrangian occur in (\ref{cL minus cR}) through $\zeta$. We are not able to fix this puzzle.\\
The main non trivial check is the fact that $c_L$ and $c_R$ in (\ref{cL and cR}) contain the results already proposed for these quantities in TMG and NMG separately.
In order to show this, we find it convenient to rewrite (\ref{cL and cR}) in a form that clearly exposes the contributions from the EH term, the NMG term and the CS term respectively. It is given by
\begin{equation}
\label{cL and cR expanded}
c_L \,=\, \frac{3\ell}{G\,\eta^2\zeta}
\left(1+\left[\frac{2}{\bar{m}^2}-1\right]-\frac{1}{3}\right)
\hspace{.7cm}
c_R \,=\, \frac{3\ell}{G\,\eta^2\zeta}
\left(1+\left[\frac{2}{\bar{m}^2}-1\right]+\frac{2\eta^2-1}{3}\right)
\end{equation}
where the square brackets isolate the contribution of the higher derivative term of NMG and the 1 just before them is due to the EH term.
From the expressions (\ref{cL and cR expanded}) we can check that, by canceling  the NMG term contained between the square brackets we obtain
\begin{equation}
\label{cL and cR TMG}
c_L \,=\, \frac{2\ell}{G\,\eta^2\zeta}
\hspace{2cm}
c_R \,=\, \frac{2\ell(1+\eta^2)}{G\,\eta^2\zeta}
\end{equation}
which is the result of \cite{Anninos:2008fx} for TMG, after having used that $\eta^2=(\nu^2+3)/(4\nu^2)$ and $\zeta=2\nu$.\\
As for the NMG regime, by canceling the contribution of the CS term in (\ref{cL and cR expanded}) (the ones after the square brackets), we get $c_L=c_R$ and in particular
\begin{equation}
\label{cL and cR nmg}
c_L \,=\, c_R \,=\,\frac{3\ell}{G\,\eta^2\zeta}\;\frac{2}{\bar{m}^2}
\,=\,\frac{6\ell\,\zeta}{G\,\eta^2 m^2}
\end{equation}
where $\eta^2$ and $\zeta$ are given by (\ref{eta2 nmg}) and (\ref{zeta nmg}) respectively (we recall that the plus sign must be taken in (\ref{zeta nmg})).
Plugging these expressions into (\ref{cL and cR nmg}) and adapting the notation as explained in the subsection \ref{section known cases}, we recover the central charge for the warped black hole in NMG proposed in \cite{Kim:2009jm} through a different method (entropy function method \cite{Sen:2005wa}).\\
We recall that the $c$ extremization method \cite{Kraus:2005vz, Saida:1999ec} that has been used to find the central charge in NMG for $AdS_3$ \cite{Bergshoeff:2009aq,Liu:2009pha}, cannot be applied in the warped case because it requires the vanishing  of terms like $D_\mu R_{\alpha\beta}$, which does not happen for the warped metrics (\ref{spacelike warped}) and (\ref{ADM gmg}).
As for (\ref{cL minus cR}), the discrepancy with the expectations from \cite{Kraus:2005vz,Kraus:2005zm} could be related\footnote{We are grateful to Per Kraus for an useful correspondence about this point.} to the fact that a complete analysis based on the boundary stress tensor is still missing for GMG (the results of \cite{Grumiller:2008qz, Skenderis:2009nt} in TMG and of \cite{Hohm:2010jc} in NMG could be helpful in this direction), or to the symmetry of the vacuum, which is not $SL(2,\mathbb{R})\times SL(2,\mathbb{R})$, or to the non unitarity of the CFT on the boundary. Concerning the last point, it is important to recall that for TMG at the chiral point the dual CFT is believed to be a Logarithmic CFT  \cite{Grumiller:2008qz,Grumiller:2009mw}, which is non unitary, and there are evidences for this in NMG as well \cite{Grumiller:2009sn,Alishahiha:2010bw}.

\section*{Conclusions}

In this paper we have studied warped black hole solutions of GMG.
A new feature occurring in this generalized model with respect to the two regimes of TMG and NMG, where these solutions are already known, is that there are regions of the parameter space providing two regular warped black holes for fixed values of the parameters.\\
A result that deserves further checks is given by the expressions (\ref{cL and cR}) of the central charges.
Despite the fact that they contain the previous proposals for TMG \cite{Anninos:2008fx} and NMG \cite{Kim:2009jm}, it turns out that $c_L -c_R$ depends not only on the coupling to the CS term, but also on the coupling to the NMG term and on the cosmological constant.\\ 
We conclude by mentioning some possible ways to go beyond this work.
The generalization to non-Einstein backgrounds of the ADT technique in NMG is needed, in order to make the first law of thermodynamics a consistency check of the procedure. 
Furthermore, recovering $c_L$ and $c_R$ through the asymptotic symmetry group analysis or through other methods would provide an important check for these expressions.
Other interesting directions to understand better the warped black holes found here could be to study their stability or the corresponding real time Green functions 
(see respectively \cite{Anninos:2009zi} and \cite{Chen:2009hg,Chen:2009cg} for these analysis in the context of TMG).

\subsection*{Acknowledgments}

It is a pleasure to thank Olaf Hohm for collaboration at the initial stage of this project, interesting discussions during its development and comments on the draft. 
I am also grateful to Dionysios Anninos, Bin Chen, Gaston Giribet, Roman Jackiw, Per Kraus and Andrew Strominger for useful discussions and correspondence.\\
This work is mainly supported by Istituto Nazionale di Fisica Nucleare (INFN) through a Bruno Rossi fellowship and also by funds of the U.S. Department of Energy (DoE) under the cooperative research agreement DE-FG02-05ER41360.

%%%%%%%%%%%%%%%%%%%%%%%%%%%%%%%%%%%%%%%%%%%%%%%%
\appendix

\section{$AdS_3$ limit of the e.o.m.}
\label{AdS3 appendix}

\noindent
In this appendix we slightly expand the discussion on the e.o.m. for the metric of warped $AdS_3$ in order to show how the relations (\ref{3 eqs system}) have been obtained for warped $AdS_3$ and how to recover a previous result on $AdS_3$ in GMG found in \cite{Bergshoeff:2009hq}.

The metric (\ref{AdS3 spacelike}) of $AdS_3$ corresponds to (\ref{spacelike warped}) with $\eta^2=1$ and $\zeta=2$, given the relations (\ref{clement subs}). We remark that it is not correct to plug these values into the expressions of the section \ref{algebraic relations} because they have been found for $\eta^2 \neq 1$. Doing it would lead to an unnecessary constraint among the parameters of the Lagrangian.\\
Plugging the metric (\ref{spacelike warped}) into (\ref{eom bis}), we find for the l.h.s. of (\ref{eom bis}) the following matrix
\begin{equation}
\label{eom matrix}
\left(\begin{array}{ccc}
\displaystyle \frac{\tilde{\sigma}}{128 \zeta^2 \eta^4}
\Big[C_3-(1-\eta^2)C_4 \cosh(2 \sigma)\Big] 
& \hspace{.7cm} 0 \hspace{.7cm} & 
\displaystyle -\frac{\tilde{\sigma} C_2}{64 \zeta^2 \eta^4}\,\sinh \sigma
\\
0 & \displaystyle \frac{\tilde{\sigma} C_1}{64 \zeta^2 \eta^2} & 0
\\
\displaystyle -\frac{\tilde{\sigma} C_2}{64 \zeta^2 \eta^4}\,\sinh \sigma
& 0 & 
\displaystyle -\frac{\tilde{\sigma} C_2}{64 \zeta^2 \eta^4}
\end{array}
\right)
\end{equation}
where (\ref{clement subs}) has been used and the $C$'s are functions of the parameters $(\eta, |\zeta|)$ of the solution and of the parameters $(\lambda_0, \alpha_0, \beta_0)$ that we find useless to report here. In particular the relevant feature of the $C$'s is their independence of the coordinate $\sigma$. Then, one notices that
\begin{equation}
C_1 \,=\,\frac{C_2-C_3}{\eta^2}\;.
\end{equation}
From (\ref{eom matrix}) it is evident the difference between $AdS_3$ and its warped version in terms of the algebraic equations to impose. Indeed, in the case of warped $AdS_3$ we have $\eta^2 \neq 1$ and we must require $(C_2=0, C_3=0, C_4=0)$ in order to satisfy the e.o.m.; while for $AdS_3$, which has $\eta^2=1$, the term with $\cosh(2\sigma)$ in the element $(1,1)$ of (\ref{eom matrix}) is not there, therefore we have to impose just $(C_2\big|_{\eta^2=1, \zeta=2}=0, C_3\big|_{\eta^2=1, \zeta=2}=0)$.
In the warped $AdS_3$ case, we can simplify the analysis by observing that
\begin{equation}
C_4 \,=\,C_2 -8\eta^2 \zeta^2 C_5\;.
\end{equation}
Thus, imposing the e.o.m. is equivalent to require $(C_2=0, C_3=0, C_5=0)$. The explicit expressions for $C_2$, $C_3$ and $C_5$ are given by the l.h.s. of the first, second and third equation of (\ref{3 eqs system}) respectively. \\
In the case of $AdS_3$ ($\eta^2=1$ and $\zeta=2$) and the conditions $(C_2\big|_{\eta^2=1, \zeta=2}=0, C_3\big|_{\eta^2=1, \zeta=2}=0)$ reduce to the same equation
\begin{equation}
\label{ads3 alpha0}
\lambda_0\,=\,\frac{\alpha_0}{4}-1
\end{equation}
which does not contain $\beta_0$, as expected, since the Cotton tensor (\ref{cotton tensor}) vanishes for $AdS_3$.\\
In \cite{Bergshoeff:2009hq,Bergshoeff:2009aq},  the maximally symmetric vacuum has been considered as a solution of the general massive gravity (\ref{gmg action}). It is defined by 
\begin{equation}
\label{max symm def}
G_{\mu\nu}\,=\,-\,\Lambda g_{\mu\nu}
\end{equation}
where $G_{\mu\nu}$ is the Einstein tensor and $\Lambda$ is a constant ($\Lambda>0$ for the de Sitter spacetime and $\Lambda<0$ for the anti-de Sitter one).  A maximally symmetric  metric is a solution of (\ref{eom bis}) if and only if $\Lambda$ solves the quadratic equation \cite{Bergshoeff:2009aq}
\begin{equation}
\label{eq max symm vacua}
\Lambda^2+4\tilde{\sigma} m^2 \Lambda-4\lambda m^4\,=\,0\;.
\end{equation}
Now, from (\ref{R warped}) with $\chi^2=\gamma^2=1$ we see that for $AdS_3$ we have $R=-6/\ell^2$ and from (\ref{max symm def}) we easily find that for a maximally symmetric vacuum $R=6\Lambda$; therefore for  $AdS_3$ we have that $\Lambda=-1/\ell^2$. Plugging this value for $\Lambda$ into (\ref{eq max symm vacua}) and using the definitions (\ref{0 parameters}), we can check that (\ref{eq max symm vacua}) reduces to (\ref{ads3 alpha0}) (see also eq. (3.9) of \cite{Clement:2009gq}).

\section{Roots of a cubic and a quartic equation}
\label{appendix roots}

In this appendix we briefly discuss for completeness the formulas for the roots of a cubic (section \ref{app cubic eq}) and a quartic equation (section \ref{app quartic eq}) that we have largely employed in the section \ref{section GMG}. For a detailed analysis we refer to the standard algebra textbooks (see e.g. \cite{Birkhoff1997,Irving2004}).
In the case of the quartic equation we briefly describe also the method because the main
equation (\ref{eq zeta}) we deal with is quartic and this discussion helps us to identify interesting special cases  where the formulas for the roots simplify.

\subsection{Cubic equation}
\label{app cubic eq}

A cubic equation with real coefficients
\begin{equation}
\label{cubic eq}
\tilde{A} \zeta^3 + \tilde{B} \zeta^2 + \tilde{C} \zeta +\tilde{D} \,=\,0
\end{equation}
has at least a real root. The occurrence of two or three distinct real roots depends on the positivity of the following quantity
\begin{equation}
\label{Delta3 def}
\Delta_3\,\equiv\,
18 \tilde{A}  \tilde{B} \tilde{C}\tilde{D}
-4  \tilde{B}^3\tilde{D}+  \tilde{B}^2\tilde{C}^2-4  \tilde{A}\tilde{C}^3
-27 \tilde{A}^2  \tilde{D}^2\;.
\end{equation}
In particular, when $\Delta_3>0$ there are three distinct real roots; when $\Delta_3=0$ the roots are all real but one of them is double and when $\Delta_3<0$ the equation has one real root and two complex conjugate ones.\\
In order to give the explicit expressions of the roots, one introduces
\begin{equation}
F\,\equiv\,\sqrt[3]{\frac{P-Q}{2}}
\end{equation}
where
\begin{equation}
P\,\equiv\,
2\tilde{B}^3
-9 \tilde{A}  \tilde{B} \tilde{C}
+27 \tilde{A}^2  \tilde{D} 
\hspace{2cm}
Q\,\equiv\, \sqrt{-\,27  \tilde{A}^2 \Delta_3}\;.
\end{equation}
Then, the three roots of (\ref{cubic eq}) read
\begin{eqnarray}
\label{rr1}
\zeta_1 &=& -\,\frac{ \tilde{B}}{3 \tilde{A}} +\frac{F}{3 \tilde{A}}
+\frac{\tilde{B}^2-3 \tilde{A} \tilde{C}}{ 3\tilde{A} F}\\
\rule{0pt}{.8cm}
\label{rr2}
\zeta_2 &=& -\,\frac{ \tilde{B}}{3 \tilde{A}} - e^{-i\frac{\pi}{3}}\frac{F}{3 \tilde{A}}
-e^{i\frac{\pi}{3}}\frac{\tilde{B}^2-3 \tilde{A} \tilde{C}}{ 3\tilde{A} F}\\
\rule{0pt}{.8cm}
\label{rr3}
\zeta_3 &=& -\,\frac{ \tilde{B}}{3 \tilde{A}} - e^{i\frac{\pi}{3}}\frac{F}{3 \tilde{A}}
-e^{-i\frac{\pi}{3}}\frac{\tilde{B}^2-3 \tilde{A} \tilde{C}}{ 3\tilde{A} F}\;.
\end{eqnarray}
These expressions have been employed to make the plots in the figures \ref{stat points picture} and \ref{no3roots picture} of the section \ref{section GMG}, which play a crucial role to get the results discussed in the section \ref{section GMG} about the number of solutions of $P_4(\zeta)=0$ allowed by the condition (\ref{no CTC condition}).

\subsection{Quartic equation}
\label{app quartic eq}

Let us consider the general quartic equation with real coefficients
\begin{equation}
\label{quartic eq}
A \zeta^4 + B \zeta^3 + C \zeta^2 +D \zeta +E\,=\,0\;.
\end{equation}
{\bf Quasi symmetric case.} We start by discussing the following simple special case
\begin{equation}
\label{quasi symmetric eq}
A \zeta^4 + B \zeta^3 + C \zeta^2 +\omega B \zeta +\omega^2 A\,=\,0\;.
\end{equation}
Dividing by $\zeta^2$ it is straightforward to see that it becomes a quadratic equation in $\zeta+\omega/\zeta$, whose solution is
\begin{equation}
\label{soln quasi sym}
\zeta+\frac{\omega}{\zeta}\,=\,
\frac{-B\pm \sqrt{B^2-4A(C-2\omega A)}}{2A}
\end{equation}
which is a quadratic equation itself; therefore we have four roots for $\zeta$, as expected.\\

\noindent 
{\bf General case.} Given the general quartic equation (\ref{quartic eq}), the first step to find its roots is to reduce it to a {\it depressed quartic equation}, namely a quartic equation without the term $\zeta^3$. This is done by introducing the variable $u$ as $\zeta=u-B/(4A)$, in terms of which (\ref{quartic eq}) becomes 
\begin{equation}
\label{dep quartic eq}
u^4+a u^2 +b u +c\,=\,0
\end{equation}
where the new coefficients are related to the ones of (\ref{quartic eq}) as follows
\begin{equation}
\label{abc def}
a\,=\, -\frac{3 B^2}{8 A^2}+\frac{C}{A}
\hspace{.6cm}
b\,=\,\frac{B^3}{8 A^3}-\frac{B C}{2 A^2}+\frac{D}{A}
\hspace{.6cm}
c\,=\,-\frac{3 B^4}{256 A^4}+\frac{B^2 C}{16 A^3}-\frac{B D}{4 A^2}+\frac{E}{A}\;.
\hspace{.1cm}
\end{equation}
When $b=0$ the equation (\ref{dep quartic eq}) reduces to a quadratic one in $u^2$ (also called {\it biquadratic equation}), which is easy to solve. If $c =0$, we have the solution $u=0$ and the roots of a depressed cubic equation, which will be discussed later in this appendix. \\
In the generic case of non vanishing $b$ and $c$, one follows the method to solve a depressed quartic equation given by Lodovico Ferrari, a mathematician who lived in the XVI century. First, (\ref{dep quartic eq}) can be equivalently written as
\begin{equation}
\label{dep quartic step1}
(u^2 + a + y)^2
=(a+2y)u^2-b u +y^2+2 a y +a^2-c 
\end{equation}
where the parameter $y$ has been introduced without restrictions. Now, if $y$ is such that the r.h.s. of (\ref{dep quartic step1}) is a perfect square, we can take the square root of (\ref{dep quartic step1}) (which introduces a first choice $\pm_{(1)}$) and find the following quadratic equation for $u$
\begin{equation}
u^2-\left(\pm_{(1)}\sqrt{a+2y}\,\right)u+
\left(\pm_{(1)} \frac{\beta}{2\sqrt{a+2y}}+a+y \right) =\,0\;.
\end{equation}
The solutions of this equation and the definition of $u$ given above lead to
\begin{equation}
\label{zeta soln}
\zeta\,=\,-\,\frac{B}{4A}+ \frac{1}{2}\left[
\pm_{(1)}\sqrt{a+2y}
\pm_{(2)}
\sqrt{-\left(3a+2y\pm_{(1)}\frac{2b}{\sqrt{a+2y}}\right)}\,
\,\right]\,.
\end{equation}
Thus, the solutions of the quartic equation are found by solving two subsequent quadratic equations which introduce two ambiguities on the sign indicated as $\pm_{(1)}$ and $\pm_{(2)}$ respectively, therefore leading to four solutions in general.\\
The condition to determine $y$ is that the r.h.s. of (\ref{dep quartic step1}) is a perfect square, i.e. its discriminant $b^2-4(a+2y)(y^2+2 a y +a^2-c )$ must vanish. This leads us to the following cubic equation
\begin{equation}
\label{cubic eq y}
y^3+\frac{5}{2}\, a y^2+(2a^2-c) y +\frac{1}{2}\left(a^3-a c-\frac{b^2}{4}\right)=\,0
\end{equation}
which can be solved by a method due to Scipione del Ferro and Tartaglia (published by Gerolamo Cardano in 1545). Again, one introduces a new variable $v$ as $y = v-5a/6$ in order to reduce (\ref{cubic eq y}) to the following {\it depressed cubic equation} (a cubic equation without the term $\zeta^2$)
\begin{equation}
\label{dep cubic eq y}
v^3 + p v + q\,=\,0
\end{equation}
whose coefficients are related to the ones of (\ref{cubic eq y}) as
\begin{equation}
\label{pq def}
p\,=\,-\frac{a^2}{12}-c
\hspace{2cm}
q\,=\,-\frac{a^3}{108}+\frac{a c}{3}-\frac{b^2}{8}\;.
\end{equation}
In our case, from (\ref{a GMG}), (\ref{b GMG}) and (\ref{c GMG}), we have
\begin{eqnarray}
\label{pq GMG}
p &=& \frac{4}{441}\left(
-\frac{132}{\alpha_0^2}+\frac{40 \beta_0^2}{\alpha_0^3}
-\frac{3\beta_0^4}{\alpha_0^4}-\frac{84\lambda_0}{\alpha_0}
\right)
\\
\rule{0pt}{.7cm}
q &=& \frac{16}{9261}\left(
\frac{272}{\alpha_0^3}-\frac{132\beta_0^2}{\alpha_0^4}
+\frac{20\beta_0^4}{\alpha_0^5}-\frac{\beta_0^6}{\alpha_0^6}
-\frac{44\beta_0^2 \lambda_0}{\alpha_0^3}
+\frac{336 \lambda_0}{\alpha_0^2}
\right)\,.
\end{eqnarray}
Notice that the cosmological parameter plays a key role in these expressions.\\
All the equations we are dealing with have real coefficients. In particular, also (\ref{dep cubic eq y}) has real coefficients and this guarantees us that it has at least a real root. Denoting by $P_3(v)$ the l.h.s. of (\ref{dep cubic eq y}) we have that $P_3(a/3)=-b^2/8 <0$, which tells us that there is at least one root in the variable $v$ strictly greater than $a/3$ (we are assuming $b\neq 0$). In terms of $y$, this means that there is at least a real root $y_0 > -a/2$. Choosing this root, we find that $a+2y_0 >0$ and therefore the expression (\ref{zeta soln}) is always well defined.
\\
As for the explicit solutions of (\ref{dep cubic eq y}), when $p=0$ the solution are simply $v_1=\sqrt[3]{- q}$, $v_2=e^{i 2\pi/3} \sqrt[3]{- q}$ and $v_3=e^{i 4\pi/3} \sqrt[3]{- q}$ where we have denoted by $\sqrt[3]{- q}$ the principal root (i.e. the one with the smallest phase);  and correspondingly we have $y_k=-5a/6-v_k$ ($k=1,2,3$). 
For non vanishing $p$, one sets $v= z+w$ and rewrites (\ref{dep cubic eq y}) as
\begin{equation}
z^3+w^3+(3z w+p)(z+w)+q\,=\,0
\end{equation}
which is solved by imposing 
\begin{equation}
\left\{\begin{array}{l}
3z w +p \,=\,0\\
w^3 +z^3 +q\,=\,0\;.
\end{array}
\right.
\end{equation}
Since $p \neq 0$, we can find $w=-p /(3z)$ from the first equation and then substituting it into the second one, which becomes a quadratic equation in $z^3$, whose solutions read
\begin{equation}
\label{z soln}
z_{\pm}^3\,=\,-\frac{q}{2}\pm\sqrt{\frac{q^2}{4}+\frac{p^3}{27}}\;.
\end{equation}
Taking the cubic root provides six solutions $\{ z_{+,k}\, , z_{-,k} \, ; k=1,2,3\}$ but one can check that the combinations $-p/(3 z_{+,k})+z_{+,k}$ and $-p/(3 z_{-,k})+z_{-,k}$ give the same three solutions.
Thus, when $p \neq 0$ we have that
\begin{equation}
\label{y soln}
y_k\,=\,-\frac{5 a}{6}-\frac{p}{3\, e^{i2(k-1)\pi/3} z_+ }+ e^{i2(k-1)\pi/3} z_+ 
\hspace{1.3cm}
k\,=\,1,2,3
\end{equation}
where again we have denoted by $z_+$ the principal root among the solutions found by taking the cubic root of the r.h.s. of (\ref{z soln}). The same $y_k$ are obtained by using $z_-$ instead of $z_+$ in (\ref{y soln}).

\section{Another useful system of coordinates}
\label{Clement coords}

The warped black hole metric has been extensively studied in \cite{Moussa:2003fc,Bouchareb:2007yx, Moussa:2008sj,Clement:2009gq}. 
In this appendix we recall the results found there in their notation and combine them for the case of the GMG. In the coordinates system used in 
\cite{Moussa:2003fc, Bouchareb:2007yx,Moussa:2008sj,Clement:2009gq} the metric of the spacelike warped black hole reads (they set $\ell =1$)
\begin{equation} 
\label{clement metric}
\frac{ds^2}{\ell^2}\,=\,
-\,N^2_{\textrm{\tiny{C}}}(\rho)\,dt_{\textrm{\tiny{C}}}^2
+\frac{d\rho^2}{\zeta^2\, r_{\textrm{\tiny{C}}}^2(\rho) N^2_{\textrm{\tiny{C}}}(\rho)}
+ r_{\textrm{\tiny{C}}}^2(\rho)\big[d\phi+N^\phi(\rho)\,dt_{\textrm{\tiny{C}}}\big]^2
\end{equation}
where
\begin{equation}
\label{rC def}
r_{\textrm{\tiny{C}}}^2(\rho)\,=\,
\rho^2+2\omega \rho+(1-\eta^2)\omega^2+\frac{\eta^2\rho_0^2}{1-\eta^2}
\end{equation}
and 
\begin{equation}
N^2_{\textrm{\tiny{C}}}(\rho)\,\equiv\,
\frac{\eta^2(\rho^2-\rho_0^2)}{r_{\textrm{\tiny{C}}}^2(\rho)}\
\hspace{1.5cm}
N^\phi(\rho)\,\equiv\,
\frac{\rho+(1-\eta^2)\omega}{r_{\textrm{\tiny{C}}}^2(\rho)}\;.
\end{equation}
This is a black hole with two horizons at $\rho = \pm \rho_0$ when $\rho_0^2 >0$. It reduces to (\ref{ADM gmg}) after the following change of coordinates
\begin{equation}
\label{coords change}
t_{\textrm{\tiny{C}}} = \frac{t}{\sqrt{1-\eta^2}}
\hspace{1.4cm}
\phi = \frac{\sqrt{\zeta^2(1-\eta^2)}}{2}\,\theta
\hspace{1.4cm}
\rho = r-\frac{r_++r_-}{2}
\end{equation}
supported by the following relations between the parameters $(r_+, r_-)$ and $(\rho_0, \omega)$
\begin{equation}
\label{rho0 omega}
\rho_0^2 \,= \left(\frac{r_+ - r_-}{2}\right)^2
\hspace{2cm}
\omega\,=\,\frac{1}{2(1-\eta^2)}\Big(r_++r_- + 2\eta \sqrt{r_+ r_-}\,\Big) \;.
\end{equation}
From (\ref{rC def}) and (\ref{rho0 omega}) one finds that
\begin{equation}
r_{\textrm{\tiny{C}}}(\rho_0)\,=\,\frac{r_+ +\eta\sqrt{r_+ r_-}}{\sqrt{1-\eta^2}}\;.
\end{equation}
The temperature and the angular velocity of the horizon can be read from (\ref{clement metric})
\begin{eqnarray}
\label{Temp clement}
T_{\textrm{\tiny{C}}} &=& \frac{1}{2\pi}\, \sqrt{g^{\rho\rho}}\,\partial_\rho N_{\textrm{\tiny{C}}}\big|_{\rho=\rho_0}
\;=\;\frac{\zeta}{4\pi\ell}\, 
r_{\textrm{\tiny{C}}}(\rho_0)\, \partial_\rho N^2_{\textrm{\tiny{C}}} \big|_{\rho=\rho_0}
\;=\;
\frac{\zeta \, \eta^2 \rho_0}{2\pi\ell \,r_{\textrm{\tiny{C}}}(\rho_0)}\\
\rule{0pt}{.7cm}
\label{Omega clement}
\Omega_{\textrm{\tiny{C}}} &=& \frac{1}{\ell}\, N^\phi  \big|_{\rho=\rho_0}
 \;=\;\frac{\sqrt{1-\eta^2}}{\ell \,r_{\textrm{\tiny{C}}}(\rho_0)}\;.
\end{eqnarray}
In order to get the entropy, the mass and the angular momentum of the warped black hole (\ref{clement metric}) in GMG, we simply sum the contributions of the EH term, the NMG term \cite{Clement:2009gq} and the CS term \cite{Moussa:2003fc, Bouchareb:2007yx,Moussa:2008sj}. The entropy is given by
\begin{eqnarray}
S_{\textrm{\tiny{C}}} &=& 
\frac{2\pi\ell\,r_{\textrm{\tiny{C}}}(\rho_0)}{4G}\left(\frac{2}{\bar{m}^2}
+\frac{r_{\textrm{\tiny{C}}}(\rho_0)^2}{3}\,  \partial_\rho N^\phi  \big|_{\rho=\rho_0}\right)
\\
\label{S clement}
\rule{0pt}{.8cm}
&=&
\frac{2\pi\ell\,r_{\textrm{\tiny{C}}}(\rho_0)}{4G}\left(\frac{2}{\bar{m}^2}
-\frac{(1-2\eta^2)\rho_0+(1-\eta^2)\omega}{3\sqrt{1-\eta^2}\,r_{\textrm{\tiny{C}}}(\rho_0)}\right)
\end{eqnarray}
where $\bar{m}^2 \equiv m^2 /\zeta^2$ has been introduced in  \cite{Clement:2009gq}. 
For the mass and the angular momentum one gets
\begin{eqnarray}
\label{M clement}
M_{\textrm{\tiny{C}}} &=& 
\frac{1}{4G}\left( \frac{2}{\bar{m}^2}-\frac{1}{3} \right)\zeta\,\eta^2(1-\eta^2)\omega
\\
\label{J clement}
\rule{0pt}{.7cm}
J_{\textrm{\tiny{C}}} &=&
\frac{\ell}{4G}\left[\left( \frac{2}{\bar{m}^2}-\frac{1}{3} \right)
\left((1-\eta^2)\omega^2-\frac{\rho_0^2}{1-\eta^2}\right)
-\frac{2\eta^2 \rho_0^2}{3(1-\eta^2)} \,\right]\frac{\zeta\,\eta^2}{2}\;.
\hspace{1cm}
\end{eqnarray}
Given these expressions, the first law of thermodynamics
\begin{equation}
dM_{\textrm{\tiny{C}}}\,=\,T_{\textrm{\tiny{C}}} dS_{\textrm{\tiny{C}}} + \Omega_{\textrm{\tiny{C}}} dJ_{\textrm{\tiny{C}}}
\end{equation}
is satisfied for independent variations of the parameters $\rho_0$ and $\omega$.\\
As emphasized in \cite{Clement:2009gq} and recalled both in the introduction and in the section \ref{entropy section}, while the mass and the angular momentum of the warped black holes in TMG have been computed through a ADT type computation \cite{Bouchareb:2007yx} and the first law of thermodynamics has been subsequently employed as a crucial test of the final expressions, in the NMG case \cite{Clement:2009gq} the super angular momentum method has been used and, in the computation of the mass, an unknown term has been fixed by imposing the first law.
Thus, for the warped black holes in NMG the first law is already implemented in the expression of the mass and the angular momentum, and it cannot be employed to check the results. Since we have used these results from NMG, it is evident that this conclusion holds for the warped black holes in GMG as well. An ADT type computation of the mass of the warped black hole in NMG is needed.\\
From (\ref{coords change}) and (\ref{rho0 omega}) one can relate the functions occurring in the metrics (\ref{ADM gmg}) and (\ref{clement metric}) expressed in the two coordinate systems (e.g. $N^\theta(r)=2N^\phi(\rho)/[\zeta(1-\eta^2)]$) and then obtain the temperature and the angular velocity of the horizon in the Schwarzschild coordinates. They read
\begin{equation}
T\,=\, \frac{1}{\sqrt{1-\eta^2}}\,T_{\textrm{\tiny{C}}} 
\hspace{2.5cm}
\Omega\,=\, \frac{2}{\zeta(1-\eta^2)}\,\Omega_{\textrm{\tiny{C}}} 
\end{equation}
where $T_{\textrm{\tiny{C}}} $ and $\Omega_{\textrm{\tiny{C}}} $ are given in (\ref{Temp clement}) and  (\ref{Omega clement}) respectively.
As for the entropy (\ref{S clement}), the mass (\ref{M clement}) and angular momentum (\ref{J clement}), we find
\begin{equation}
\label{SMJ change}
S \,=\, \frac{\zeta\sqrt{1-\eta^2}}{2}\,S_{\textrm{\tiny{C}}} 
\hspace{2cm}
M \,=\, \frac{\zeta}{2}\,M_{\textrm{\tiny{C}}} 
\hspace{2cm}
J\,=\, \frac{\zeta^2(1-\eta^2)}{4}\,J_{\textrm{\tiny{C}}} 
\end{equation}
in the Schwarzschild coordinates, which are the expressions given in (\ref{Ss warped}), (\ref{Ms warped}) and (\ref{Js warped}) respectively.

\section{Self-dual and null z-warped black holes}
\label{appendix selfdual and null}

In this appendix we find that also self-dual warped black holes (section \ref{app selfdual}) and null z-warped black holes (section \ref{app null}) occur in GMG. These solutions have been already studied 
in TMG in \cite{Chen:2010qm}\footnote{We are grateful to Bin Chen for having drawn our attention to \cite{Chen:2010qm}.} and \cite{Gibbons:2008vi,Anninos:2010pm} respectively. Here we briefly give the solutions, leaving a deeper analysis of them to future work.

\subsection{Self-dual warped black holes}
\label{app selfdual}

Let us consider the following metric in the coordinates $\{\tilde{\tau}, x, \psi\}$, which is a slight modification of the one studied in \cite{Chen:2010qm} for TMG (see also \cite{Coussaert:1994tu} for the EH analogues)
\begin{eqnarray}
\label{selfdual warped}
ds^2  &=& \frac{\ell^2 \chi^2}{4}\,
\bigg\{ -(x-x_+)(x-x_-)d\tilde{\tau}^2
+\;\frac{dx^2}{(x-x_+)(x-x_-)}\\
\rule{0pt}{.75cm}
& &\hspace{6.7cm}
+\;\gamma^2
\left[\,\xi\,d\psi+\left(x-\frac{x_+ +x_-}{2}\right)d\tilde{\tau}\,\right]^2
\,\bigg\}\;.\nonumber
\end{eqnarray}
As for the coordinates, we have $\tilde{\tau} \in \mathbb{R}$, $x \in \mathbb{R}$ and $\psi \sim \psi +2\pi$.
The Ricci scalar of (\ref{selfdual warped}) is $R=-6/\ell^2$ and its determinant reads $g=-\ell^6\gamma^2\chi^6/64$. The vacuum solution corresponds to $x_+ =x_- =0$ but it is not clear
the corresponding value of the parameter $\xi$ \cite{Chen:2010qm}. \\
Plugging this metric into the e.o.m. (\ref{eom bis}) and employing the parameterization (\ref{clement subs}), after some algebra one recovers the algebraic relations (\ref{3 eqs system}).
This is expected because the metric (\ref{selfdual warped}) is locally related to the spacelike warped black hole (\ref{ADM gmg}) through the following linear coordinates transformation
\begin{eqnarray}
t &=&\frac{1}{\eta^2\zeta}
\left[\,\xi\,\psi-\left(\frac{r_+ + r_-}{2}+ \eta \sqrt{r_+ r_-}\right)
\frac{x_+ -x_-}{r_+ - r_-}\,\tilde{\tau}\,\right]\\
\rule{0pt}{.8cm}
\theta &=& \frac{2}{\eta^2\zeta^2}\; \frac{x_+ -x_-}{r_+ - r_-}\,\tilde{\tau}\\
\rule{0pt}{.8cm}
r &=& \frac{r_+ - r_-}{x_+ - x_-}\left(x-\frac{x_+ + x_-}{2}\right)+\frac{r_+ + r_-}{2}\;.
\end{eqnarray}
Thus, (\ref{selfdual warped}) is locally equivalent to the spacelike warped $AdS_3$ (\ref{AdS3 spacelike}) as well.\\
We recall that  the metric of the form (\ref{selfdual warped}) has been studied in TMG \cite{Chen:2010qm} because it resembles the one obtained by fixing the polar angle in the near horizon geometry of the near extremal Kerr black hole (near NHEK) \cite{Bredberg:2009pv}.

\subsection{Null z-warped black holes}
\label{app null}

We consider the following metric in the coordinates $\{\tilde{t},\tilde{\theta}, \tilde{r}\}$ \cite{Anninos:2010pm} 
\begin{equation}
\label{null warped}
\frac{ds^2}{\ell^2}\,=\,
2\tilde{r}\, d\tilde{t} d\tilde{\theta}+
\big[\,\tilde{r}^z+b \tilde{r}+\delta \tilde{r}\,\tilde{t}+a(\tilde{\phi})^2\,\big]d\tilde{\theta}^2
+\frac{d\tilde{r}^2}{4\tilde{r}^2}
\end{equation}
where $\tilde{\theta}$ is a periodic coordinate, $\tilde{\theta}\sim \tilde{\theta} +2\pi$.
The parameters $b$ and $\delta$ are constant while $a(\tilde{\phi})$ is an arbitrary periodic function which characterizes a hairy black holes when it is not a constant. The Ricci scalar of (\ref{null warped}) is $R=-6/\ell^2$. 
Plugging (\ref{null warped}) into the e.o.m. (\ref{eom bis}) we find that the first relation to impose is (\ref{ads3 alpha0}), namely the same relation obtained for $AdS_3$. Given (\ref{ads3 alpha0}), the metric (\ref{null warped}) is a solution of GMG for either $z=0$ or $z=1$ or the roots of the following quadratic equation
\begin{equation}
\label{null z eq}
16(1+\lambda_0)z^2-2(8+\beta_0+8\lambda_0)z+3+\beta_0+2\lambda_0
\,=\,0
\end{equation}
namely
\begin{equation}
\label{z null warped}
z\,=\,\frac{1}{2}+\frac{\beta_0\pm\sqrt{\beta_0^2+16(1+\lambda_0)(1+2\lambda_0)}}{16(1+\lambda_0)}\;.
\end{equation}
In the special case of $\beta_0=0$ this analysis provides solutions of NMG. The result of \cite{Anninos:2010pm} in TMG is recovered by setting $\lambda_0=-1$ (this tells us that we are in TMG through (\ref{ads3 alpha0})) in (\ref{null z eq}), which simplifies to a first order equation whose solution is $z=(1/\beta_0+1)/2$.\\
As emphasized in \cite{Anninos:2010pm}, the near horizon geometry of (\ref{null warped}) with constant $a(\tilde{\phi})$ and $\delta=0$ is the self-dual orbifold of $AdS_3$ \cite{Coussaert:1994tu}.
A detailed study of the metrics (\ref{null warped}) in GMG along the lines of \cite{Anninos:2010pm} is needed and we leave it to future work.

%%%%%%%%%%%%%%%%%%%%%%%%%%%%%%%%%%%%%%%%%%%%%%%%%%%

\end{document}